\newif\ifShowKeys
\ShowKeysfalse

\newif\ifLocalFigs
\LocalFigstrue

\documentclass[11pt,a4paper]{article} 
\pdfoutput=1
\usepackage[no-natbib-sort]{my-jheppub}

\usepackage{amsmath, amssymb}

\usepackage{mathpazo}
\usepackage{bm}
\usepackage{environ}
\usepackage{mathrsfs}
\usepackage{array,arydshln}

\usepackage{graphicx,epsfig}
\usepackage{epic}

\usepackage{tensor} 							

\usepackage{float}
\usepackage[dvipsnames]{xcolor}
\definecolor{maroon}{rgb}{0.8,0.3,0.}

\usepackage{slashed}
\usepackage[nodayofweek]{date time}

\ifShowKeys \usepackage[notcite]{showkeys} \fi

\usepackage{hyperref}

\usepackage{aurical}
\usepackage[T1]{fontenc}

\usepackage{framed}
\definecolor{shadecolor}{RGB}{255, 230, 204}

\allowdisplaybreaks

\newmuskip\pFqmuskip

\newcommand*\pFq[6][8]{%
  \begingroup 
  \pFqmuskip=#1mu\relax
  \mathcode`\,=\string"8000
  \begingroup\lccode`\~=`\,
  \lowercase{\endgroup\let~}\pFqcomma
  {}_{#2}F_{#3}{\left[\genfrac..{0pt}{}{#4}{#5};#6\right]}%
  \endgroup
}

\newcommand*\pFtildeq[6][8]{%
  \begingroup 
  \pFqmuskip=#1mu\relax
  \mathcode`\,=\string"8000
  \begingroup\lccode`\~=`\,
  \lowercase{\endgroup\let~}\pFqcomma
  {}_{#2}\widetilde{F}_{#3}{\left[\genfrac..{0pt}{}{#4}{#5};#6\right]}%
  \endgroup
}

\newcommand{\pFqcomma}{\mskip\pFqmuskip}

\newcommand{\be}{\begin{equation}}
\newcommand{\ee}{\end{equation}}

\newcommand{\rf}[1]{(\ref{#1})}

\newcommand{\mc}{\mathcal }

\newcommand{\bv}{\begin{verbatim}}
\newcommand{\ev}{\end{verbatim}}

\newcommand{\la}{\label}

\newcommand{\eps}{\varepsilon}
\newcommand{\wt}{\widetilde}

\newcommand{\vp}{\varphi}
\newcommand{\ovp}{\overline\varphi}

\title{On induced action for conformal higher spins \\    in curved   background }
\author[a]{Matteo Beccaria} 
\author[b]{ and \ \ Arkady A. Tseytlin\footnote{Also at Lebedev Institute, Moscow.}}

\abstract{
We continue the investigation of the structure of the  action for a tower of conformal  higher spin  fields 
in non-trivial  4d background metric  recently  discussed  in 
\href{http://arxiv.org/abs/1609.09381}{arXiv:1609.09381}.
The action is defined   as an induced one  from   path integral of a  conformal scalar  field in curved background 
   coupled  to   higher spin fields. We analyse in detail the dependence  of the quadratic  part of the induced action 
on the spin 1  and spin 3  fields, determining the presence of a  curvature-dependent 
mixed  spin  1--3 term.  One consequence    is that   the pure spin 3 kinetic term  cannot be gauge-invariant 
on its own beyond the  leading term  in  small curvature expansion. 
We  also compute the  non-zero contribution   of the  1--3 mixing term  to the  conformal anomaly c-coefficient. 
One is thus to  determine  all such mixing   terms before 
addressing the question   of    possible  vanishing of the total c-coefficient in the  conformal higher spin theory. 
}

\affiliation[a]{Dipartimento di Matematica e Fisica Ennio De Giorgi,\\
Universit\`a del Salento \& INFN, Via Arnesano, 73100 Lecce, 
Italy} 

\affiliation[b]{The Blackett Laboratory, Imperial College, London SW7 2AZ, U.K.}
                       
\emailAdd{matteo.beccaria@le.infn.it} \emailAdd{tseytlin@imperial.ac.uk}

\begin{document}


 \begin{flushright}\small{Imperial-TP-AT-2017-{02}}\end{flushright}				

\maketitle
\flushbottom

 \def \TT  {{\rm T}}
\def \OO {{\mc O}} 
\def \te {\textstyle} \def \ci {\cite}\def \ov {\over}  \def \ha {\te {1\ov 2}}
\def \third {\te {1\ov 3}}\def \foot {\footnote}
\def \iffa {\iffalse } 
\def \del {\partial}
\def \na {\nabla}
\def \six { {\te {1 \ov 6}} }
\def \intt {{\rm int}}
\def \td {\tilde}
   \def \SS {{\rm S}} 
\def \G {\Gamma}

\def \hA {\widetilde h} \def \hF {\widetilde F} 
\def \rc  {{\rm c}} 
\def \dd {{\rm d}}
\def \bea  {\begin{align}}  \def   \eea  {\end{align}}
 \def \TT  {{\rm T}}

\def \no {\notag} \def \ed {\end{document}}

\section{Introduction}

Conformal higher spin (CHS) theories \cite{Fradkin:1985am,Fradkin:1989md}  
in 4 dimensions generalize the Maxwell  (${s=1}$) and Weyl ($s=2$)  theories  to higher rank totally symmetric tensors 
$h_{(s)} = (h_{a_1 ...a_s})$. They have a local but   higher-derivative   free 
action $\text{S}_{s} = \int d^{4}x\,h_{(s)}\,\text{P}_{s}\,\partial^{2s}\,h_{(s)}$   
with  maximal  spin $s$ gauge  symmetries,  $\delta h_{(s)} = \partial \eps_{(s-1)}+\eta_{(2)}\,\xi_{(s-2)}$, 
allowing to choose a gauge where $h_{(s)}$ is  transverse  and traceless. 

These symmetries can be systematically   described and extended  to non-linear interacting level  
 by considering the coupling 
of CHS fields  $h_{(s)} J^{(s)}$  to   conserved   currents of a  free complex  scalar theory, 
$J_{(s)}=\ovp \,\mc J_{(s)}\,\vp $, \ $\mc J_{(s)}=\partial^{s}+...$.
 Integrating out the scalar field 
one  can then obtain a   local  invariant   interacting 
 action for an  infinite tower ($s=0,1,2, ...$)  of the CHS   fields   as  an {\it  induced} action, {\em i.e.} as the 
 coefficient of the   local (logarithmically UV divergent) part of the scalar effective action 
  $\SS(h) = \log\det\big[-\partial^{2}+\sum_{s} h_{(s)}\,\mc J_{(s)}\big] \big|_{\rm UV}$
  \ci{Tseytlin:2002gz,Segal:2002gd,Bekaert:2010ky,Giombi:2013yva}.

While it is not clear   how to  write down   this  induced action to all orders  in an  explicit form, 
few   leading  cubic and quartic  interaction terms  beyond  the free $h_{(s)} \del^{2s}h_{(s)}$  term 
  can be found  by direct   diagrammatic expansion in powers of $h_{(s)}$ 
\cite{Bekaert:2010ky,Joung:2015eny,Beccaria:2016syk}. One can  then  compute  some  simplest 
4-particle scattering amplitudes   due to  the exchange of the infinite  tower of the CHS fields 
 and conclude that they vanish  \cite{Joung:2015eny,Beccaria:2016syk}  which may be attributed
  to the presence of a  global  conformal higher spin symmetry. 

To address the question of possible 
anomalies  in the quantum  CHS theory \cite{Fradkin:1985am,Giombi:2013yva,Tseytlin:2013jya}
one needs to go beyond a  perturbative near-flat-space  expansion   and  determine, {\em e.g.}, 
the  generalization of the  CHS   quadratic    terms  $h_{(s)} \del^{2s}h_{(s)}$  
to a curved background metric.   As the  free   flat-space CHS theory  is conformally invariant, 
this  is  relatively straightforward to  do  for  a homogeneous conformally-flat   background 
 ($S^{4}$, (A)dS$_{4}$, or $\mathbb R\times S^{3}$):   in this case 
the  spin $s$ CHS kinetic operator is  known  explicitly and  it  factorizes into a 
product of  $s$   second-order differential operators 
\cite{Tseytlin:2013jya,Metsaev:2007rw,Metsaev:2014iwa,Nutma:2014pua,Beccaria:2014jxa}.\foot{This  factorization 
allowed to show that the CHS theory  has vanishing 1-loop Casimir energy on $\mathbb R 
\times S^3$ \cite{Beccaria:2014jxa} as well as the  trivial partition function on $S^4$ 
    \cite{Beccaria:2015vaa}. 
    This is  consistent with the vanishing of the  1-loop  conformal anomaly a-coefficient 
    after summing up  all the spin $s$ contributions \cite{Giombi:2013yva,Tseytlin:2013jya}.} 
    

  The  case of a  general  background metric  appears to be    much more complicated.\foot{One may 
  attempt to construct a  covariant  generalization of the  CHS $\del^{2s}$   operator  by  just  imposing the required symmetries 
  (covariant analog of spin $s$ gauge invariance and Weyl symmetry).
   This approach suggests  that a  spin $s$    CHS  operator will no longer factorize  if the metric is not 
   conformally  flat  and allows  to construct  the spin 3 operator  to linear order in a 
    small curvature expansion  \ci{Nutma:2014pua}.}
    As the  conformal  spin 2  field $h_{ab}$    should be the 
fluctuating part of the metric,  $g_{ab}=\eta_{ab} + h_{ab}$,   finding the background-covariant  generalization
of the   CHS kinetic terms     is equivalent to finding 
 an infinite class of  interaction terms  in the above induced action   containing an  arbitrary power $n$ of the spin 2 field, {\em i.e.} 
    $  h_{(s)}  h_{(s')}   (h_{(2)})^n $. 

An alternative   approach
(which should be  equivalent to a
resummation of the near-flat-space expansion)  was suggested in \cite{Grigoriev:2016bzl}. 
One starts with  an  effective particle Hamiltonian in CHS background 
 that  makes  explicit the full non-linear symmetry of the theory  generalizing the construction  of 
  \cite{Segal:2002gd}. 
  Quantization of this Hamiltonian gives  a  covariant conformal scalar action  in $g_{ab}$ background 
 coupled also to   the CHS  fields.
This  determines  the  background-covariant  generalization of the symmetries 
 acting on  the scalar field  $\vp$ and 
$h_{(s)}$.  
An  underlying assumption is that  the   induced CHS action should admit the vacuum  with 
the metric $g_{ab}=\eta_{ab} + h_{ab}$   satisfying the Bach-flatness condition  (Weyl gravity  equation of motion)
with all other $h_{(s)}$ fields   being zero. It then follows that the 
resulting CHS kinetic operator  should be gauge-invariant, 
at least to the leading  order in small curvature expansion (generalizing the  $s=3$ result of \ci{Nutma:2014pua}). 

Another consequence of the   background-covariant  generalization of the  CHS symmetries suggested in 
\cite{Grigoriev:2016bzl}   is  that,  in contrast to earlier  expectations,  the 
curved  space  analog of the CHS $\del^{2s}$  kinetic  operator   should  not, in general, be diagonal in spin $s$. 
In particular, the   spin 1 and spin 3 fields    should mix via  the curvature terms   like  $R_{....} \nabla h_{(1)} \nabla h_{(3)} + ...$
 \cite{Grigoriev:2016bzl}. These  mixing terms   vanish in a  conformally flat Einstein space
  but are non-trivial in general.

The  aim  in the present paper   will be to elaborate on   the  background-covariant  approach of  \cite{Grigoriev:2016bzl}, 
{\em i.e.}   to couple the CHS  fields  to a scalar field defined  on a curved background 
and then directly compute the resulting induced action, explicitly determining the   
form of the  spin 1 -- spin 3 mixing term  anticipated in \cite{Grigoriev:2016bzl}. 
One implication  of  the  presence of this term is  that the pure spin 3 quadratic term   in  the induced action 
    cannot  be gauge invariant on its own, beyond the terms   
    linear  in the small curvature expansion    \ci{Nutma:2014pua}   (we  will   confirm the result of  \ci{Nutma:2014pua} 
    for such terms 
     directly from the    induced action  approach).\footnote{It would interesting   to  see how these conclusions   are consistent 
with the supersymmetry-based  approach of  \cite{Kuzenko:2017ujh}   where  it was suggested to couple  
 $\mc N=1$ superconformal
  higher-spin multiplets (e.g. 
containing spin 3 field)   with  $\mc N=1$   conformal supergravity (containing  conformal spins 2, 1  and 3/2).
 }

We will  also  show that this  1--3  mixing term gives a non-trivial contribution to the  UV divergences 
and hence  to the 
conformal anomalies  of the CHS theory: 
while  it does not contribute to the anomaly a-coefficient, it contributes to the  c-coefficient. 
 Thus similar mixing terms  are to be accounted for when 
 addressing the question of cancellation of  conformal c-anomaly \cite{Tseytlin:2013jya}  in the   full CHS theory.

We will start in  section \ref{sec:classical} with a  discussion  
 of the curved space   analogs of the  conserved  flat space scalar field  currents that can be 
coupled to the higher spin fields $h_{(s)}$. We will see that the  candidate spin 3 current is  conserved only modulo curvature terms 
 implying that the spin 3   coupling is not gauge invariant  by itself. This  
  non-invariance can be compensated by a non-trivial transformation of the spin 1 field as, indeed,  was  predicted    by  the 
general analysis of \cite{Grigoriev:2016bzl}. 
We will also   discuss the difference  between the "on-shell" (using scalar  field equation)
 and    "off-shell"  (manifest)   symmetries  that  require   introduction of extra couplings  nonlinear in $h_{(s)}$  
 (one   should  be able to absorb the latter  into a redefinition of the tower of CHS fields  to establish an equivalence  to 
  the approach of \cite{Grigoriev:2016bzl}). 
  These non-linear terms  may, in principle,  contribute "contact terms" to the resulting  induced action.

In section \ref{sec:ind} we  will   review the   general  structure   of the induced CHS action   in a curved space
background  starting with the well-known cases  of    
  spin $1$ and $ 2$  terms. 
  The  dependence of the quadratic part of the  induced action 
  on  the  spin 3   field  will be    studied in detail in section \ref{sec:ind-3}. 
In particular, we will compute the 1--3  mixing action  by an explicit covariant 
background field expansion. 
We will also discuss the  direct computation  of the   pure spin 3  kinetic term   to the leading order   in   the weak curvature expansion, 
finding     agreement with the    result  of \cite{Nutma:2014pua} found  from symmetry considerations.

In section \ref{sec:CC}   we will   determine  the contribution of the mixed 1--3 term to the Weyl-squared   UV divergences, {\em i.e.} to the
conformal   anomaly  c-coefficient. 

Some technical details   will appear in several Appendices. 
 In  particular, in  Appendix  \ref{app:T3}  we  will  find   the  traceless,  conserved   and   (on-shell)  gauge-invariant  
 stress tensor  for the  free   spin 3  theory  verifying  its  conformal invariance in the flat space.
   In Appendix  \ref{app:linear}  we will show  the vanishing of  the  linear in $h_{(3)}$   term 
 in the   CHS action  in an arbitrary curved  background   which is   
 consistent  with the general  expectation   of the vanishing of the linear terms in the induced action for all spins 
  in Bach-flat   backgrounds \cite{Grigoriev:2016bzl}.

\section{Scalar  field   coupled to    conformal higher   spins  }
\la{sec:classical}

Below we   will  discuss the coupling of   conformal higher spin fields  to   bilinear currents of a  complex  scalar field.
This is well a  known   story  in  flat space   \ci{Segal:2002gd,Bekaert:2010ky} 
 but the case of a  curved metric background is much more complicated 
and was addressed only recently  in \ci{Grigoriev:2016bzl}. 
Here we will use  a direct approach  based on  attempting to construct  conserved traceless  currents with correct flat-space limit.
We   will   concentrate  on  the low spin cases $s \leq 3$.

\subsection{Flat space background}

Let us  start with  a massless complex scalar in 4d flat Minkowski  space ($g_{ab}=\eta_{ab}$)  with the  action 
$S_{0} = \int d^{4}x\,\ovp\,\partial^{2}\,\vp$. 
As is well known, one   can build  in a unique way bilinear   currents 
that are traceless totally symmetric  rank $s$ tensors $J_{a_1...a_s}$  and
are  conserved on  the  scalar equations of motion $\Box \vp=0$, {\em i.e.} \cite{Craigie:1983fb,Berends:1985xx}\foot{Below  we will  sometimes use    shortcut notation:
$h_{a_{1}\cdots a_{s}}\equiv h_{(s)}$, \
$J^{a_{1}\cdots a_{s}}\equiv J^{(s)}$, \ 
$J^{(s)}\, h_{(s)} = J^{a_{1}\cdots a_{s}}\,h_{a_{1}\cdots a_{s}}$  and 
$\partial_{a_{1}\cdots a_{s}} = \partial_{a_{1}}\cdots \partial_{a_{s}}$.
Symmetrization of indices will be  the  weighted one   as  in $A_{(a} B_{b)} = {1\ov 2!} (A_a B_b + A_b B_a)$, {\em etc.}}
\be
\la{2.1}
\partial^{a_1}\, J_{a_{1}\cdots a_{s}}=0,\qquad \qquad J\indices{^{a_1}_{a_1\cdots a_{s}}}=0\ .
\ee
The  lowest-spin examples are   
\begin{align}
J_a &= i\,\ovp\,\partial_{a}\,\vp+\text{c.c.}, \qquad \qquad 
J_{ab} =  (\ovp\, \partial_{a}\partial_{b}\,\vp 
-2\ \partial_{a}\ovp\, \partial_{b}\vp 
+ c.c.) + g_{ab}\,\partial^{c}\ovp\, \partial_{c}\vp\ ,  \la{22} \\
J_{abc} &=  
6{i} 
\Big[
\ovp\,\partial_{a}\partial_{b}\partial_{c}\,\vp
-9\,\partial_{(a}\,\ovp\,\partial_{b}\partial_{c)}\,\vp
+3\,g_{(ab}\,\partial^{p}\,\ovp\,\partial_{p}\partial_{c)}\,\vp \Big]+\text{c.c.}\ . \la{2.2}
\end{align}
 The properties (\ref{2.1})
imply that the currents may be coupled to conformal  higher spin fields $h_{a_{1}\cdots a_{s}}$ by 
adding to $S_{0}$ the source term
\be
\la{2.3}
S_{\text{int}}=\sum_{s} \int d^{4}x\,h^{a_{1}\cdots a_{s}} (x) \,J_{a_{1}\cdots a_{s}}\ .
\ee
This coupling term is  then   invariant under the linearized  higher spin gauge and  the  algebraic  (or "generalised  Weyl")  transformations
\be
\la{2.4}
\delta h_{a_{1}\cdots a_{s}} = \partial_{(a_{1}}\,\eps_{a_{2}\cdots a_{s})}
+g_{(a_{1}a_{2}}\,\xi_{a_{3}\cdots a_{s})} \ , 
\ee
provided one  is allowed  to drop terms proportional to  the free  scalar field 
equation of motion. This  linearized   on-shell invariance 
can then be extended  to   an  off-shell  invariance of $S_0 + S_{\text{int}}$ 
   if  one also  transforms  the scalar field  and adds  terms linear in $h_s$  to \rf{2.4} 
    (see \ci{Segal:2002gd,Bekaert:2010ky}  for a general  discussion). 
   
One may  fix the algebraic  invariance by imposing  the   traceless condition 
$h\indices{^{a_1}_{a_1\cdots a_{s}}}=0$. The residual gauge transformations  preserving this 
condition are, {\em e.g.}, 
\begin{align}
\la{2.5}
\delta h_{ab} &=\te  \partial_{(a}\,\eps_{b)}- {\te {1\ov 4} }\,g_{ab}\,\partial^c \eps_c \ , 
 \\
 \la{2.6}
\delta h_{abc} &=\te  \partial_{(a}\,\eps_{bc)} 
-\frac{1}{3}\, g_{(ab}\,\partial^{p}\,\eps_{c)p}\ , 
\qquad   \eps_{ab}= \eps_{ba}, \ \ \ \eps\indices{^{a}_{a}}=0 \ . 
\end{align}


\subsection{Curved space   background:   spins  $s \leq 3$ }  

Switching on a curved background metric,   our   starting point   will be  the action of a conformally coupled scalar
 with a  higher source term  
\begin{align}
\la{2.7}
&S = S_{0}+S_{\text{int}}\ , \notag \\
&S_{0} =  \int d^{4}x\,\sqrt{g}\,\ovp \Big(\nabla^{2}-{ \te\frac{1}{6}}\,R\Big)\,\vp,\qquad\qquad 
S_{\text{int}} = \sum_{s} \int d^{4}x\,\sqrt{g}\,  h^{a_{1}\cdots a_{s}} (x) \,J_{a_{1}\cdots a_{s}} \ .
\end{align}
To find  suitable   higher spin currents   $J_{(s)}$
we shall   generalize \rf{2.1}  and  require the   covariant conservation of the currents
 $\nabla^{a_{1}}\,J_{a_{1}\cdots a_{s}}=0$ 
on  the scalar  equations of motion ${(\nabla^{2}-\frac{R}{6})\,\vp=0}$
and tracelessness.  We will  also  add  the condition of  local  Weyl invariance of both $S_0$ and $S_{\rm int}$ 
(that implies conformal invariance in flat limit)  under   
\be
\la{2.9}\delta_{\rm w} g_{ab}= 2\,\omega\,g_{ab}\ ,\qquad\quad  
\delta_{\rm w}\vp=-\omega\,\vp\ ,\quad \qquad 
\delta_{\rm w} h_{a_{1}\cdots a_{s}} = 2\,(s-1)\,\omega\,h_{a_{1}\cdots a_{s}}\ , 
\ee
{\em i.e.} will thus  demand 
\be
\la{2.8}
\nabla^{a_1}\, J_{a_{1}\cdots a_{s}}=0\ ,\qquad 
J\indices{^{a_1}_{a_{1}\cdots a_{s}}}=0\ ,\qquad 
\delta_{\text{w}} \int d^{4}x\,\sqrt{g}\, h^{(s)}\,J_{(s)} = 0 \ . 
\ee
In general, we will  also require  that the curved space currents $J^{(s)}$  have the standard flat space
limit  \rf{22},(\ref{2.2}).
If the  tracelessness and the covariant conservation conditions in \rf{2.8}  were   possible to satisfy we would get 
(assuming that we   may use the  scalar field  equations) 
the  covariant  generalization of the transformations \rf{2.4}, {\em i.e.} 
\be
\la{2.44}
\delta h_{a_{1}\cdots a_{s}} = \na_{(a_{1}}\,\eps_{a_{2}\cdots a_{s})}
+g_{(a_{1}a_{2}}\,\xi_{a_{3}\cdots a_{s})} \ . 
\ee
As is well known, the  three conditions in (\ref{2.8})   can be indeed   satisfied for  spins $1$ and $2$. The spin 1 case
the  current is the same as in flat space  \rf{22} and is again   conserved on-shell,  {\em i.e.}   
\be
\la{2.10}
J_{a} = i\,\big(\ovp\,\nabla_{a}\vp-\nabla_{a}\ovp\,\vp\big) \ , \ \ \ \ \ \ \ \ \ \  \nabla^a J_a =0 \ . 
\ee
The source term $ \sqrt g  g^{ab} h_a J_b$  is Weyl invariant if  $h_a$    has  weight zero,  in 
agreement with (\ref{2.9}). The most general Ansatz for the  spin 2 
current   (with correct flat limit  in  \rf{22}  for $k_1=-2, \ k_2=1$)  is 
\be
\begin{split}
\la{2.11} \!\!\!
J_{ab} = (\ovp\, \nabla_{a}\nabla_{b}\,\vp+
k_{1}\,\nabla_{a}\ovp\,\nabla_{b}\vp+\text{c.c})
+k_{2}\,g_{ab}\,\nabla_{c}\ovp\,\nabla^{c}\,\vp 
+ (k_{3}\,R_{ab}+ k_{4}\,g_{ab}\,R)   \,\ovp\,\vp .
\end{split}
\ee
Imposing the Weyl invariance of  $\sqrt{g}\,h^{ab}\,J_{ab}$ with 
$\delta_{\rm w} h_{ab} =2\,\omega\,h_{ab}$  as in   (\ref{2.9}) 
we find  that $k_{1}=-2$ and $k_{3}=-1$. The trace condition $J\indices{^{a}_{a}}=0$ gives
$k_{2}=1$ and $k_{4}=\frac{1}{6}$. With these coefficients, the current is automatically
conserved  on-shell, $\nabla^{a}\,J_{ab}=0$. 
One can then check that this  $J_{ab}$ is indeed  the  stress tensor of the conformally coupled scalar with action $S_0$ in \rf{2.7} 
\be
\la{2.12}
J_{ab} = \frac{6}{\sqrt{g}}\,\frac{\delta S_{0}}{\delta g^{ab}}
= (\ovp\, \nabla_{a}\nabla_{b}\,\vp-2 \,\nabla_{a}\ovp\,\nabla_{b}\vp+\text{c.c})
+\,g_{ab}\,\nabla_{c}\ovp\,\nabla^{c}\,\vp  -  (R_{ab}-     \six
 \,g_{ab}\,R)   \ovp\,\vp \ .
\ee
The higher spin cases $s \geq 3$ display new features. 
The most general Ansatz for the  spin 3 current on a curved  background is\footnote{Here the coefficient $k_{0}$ is introduced for generality but will be 
fixed to 1 later.}
\begin{align}
\la{2.14}
J_{abc} &=6 i 
\Big[
k_{0}\,\ovp\,\nabla_{(a}\nabla_{b}\nabla_{c)}\,\vp
+k_{1}\,\nabla_{(a}\,\ovp\,\nabla_{b}\nabla_{c)}\,\vp
+k_{2}\,g_{(ab}\,\nabla^{p}\,\ovp\,\nabla_{p}\nabla_{c)}\,\vp \notag \\
&\ \ +k_{3}\,g_{(ab}\,\ovp\,\nabla^{2}\,\nabla_{c)}\,\vp
+k_{4}\,g_{(ab}\,R\,\ovp\,\nabla_{c)}\,\vp 
+k_{5}\,R_{(ab}\,\ovp\,\nabla_{c)}\,\vp 
\Big]+\text{c.c}. \ . 
\end{align}
Imposing the trace condition $J\indices{^{a}_{ab}}=0$  gives  
\be
\la{2.15}
k_{2}=-\tfrac{1}{3}\,k_{1},\qquad k_{4}=\tfrac{1}{12}\,k_{0}+\tfrac{1}{36}\,k_{1}+\tfrac{1}{3}\,k_{3},
\qquad k_{5}=-k_{0}-3\,k_{3}\ .
\ee
Once the current is traceless  the  coupling 
\be\la{2.13}
  \int d^{4}x\,\sqrt{g}\,h^{abc}\,J_{abc} \ ,
\ee
is invariant   under the algebraic  symmetry in \rf{2.4}, {\em i.e.}  
$\delta h_{abc} = g_{(ab}\,\xi_{c)}$, allowing to fix  the traceless   gauge on $h_{abc}$, 
\be  h^a_{\ ab} =0 \ . \la{213} \ee
In this  gauge    the  $g_{ab}$ terms in \rf{2.14}   decouple  ({\em i.e.} can be dropped in \rf{2.14}) 
and then the Weyl invariance of (\ref{2.13})  under \rf{2.9} ({\em i.e.} 
$\delta_{\rm w} h_{abc} = 4\,\omega\,h_{abc}$) gives the constraints 
\be
\la{2.16}
k_{1} = -9\,k_{0}, \qquad \qquad k_{5}=-7\,k_{0}\ .
\ee
Thus, the unique traceless current  that gives a Weyl invariant source  term \rf{2.13} 
for the  traceless spin 3 field is (\ref{2.14}) with 
\be
\la{2.17}
(k_{1}, k_{2}, k_{3}, k_{4}, k_{5}) = \big(-9,3,2,\tfrac{1}{2}, -7\big)\,k_{0} \ . 
\ee
The explicit form of this $J_{abc}$  that has the right flat space limit \rf{2.2}
 is thus 
(we  set  $k_{0}=1$) 
\begin{align}
\la{2.18}
J_{abc} &= 6i 
\Big[ \ovp\,\nabla_{(a}\nabla_{b}\nabla_{c)}\,\vp
-9\,\nabla_{(a}\,\ovp\,\nabla_{b}\nabla_{c)}\,\vp
+3\,g_{(ab}\,\nabla^{p}\,\ovp\,\nabla_{p}\nabla_{c)}\,\vp \notag \\
&\ \qquad  +2\,g_{(ab}\,\ovp\,\nabla^{2}\,\nabla_{c)}\,\vp
+\tfrac{1}{2}\,g_{(ab}\,R\,\ovp\,\nabla_{c)}\,\vp 
-7\,R_{(ab}\,\ovp\,\nabla_{c)}\,\vp 
\Big]+\text{c.c}.
\end{align}
Its covariant  divergence   can be    simplified   (using  the scalar field equations of motion) to 
\begin{align}
\la{2.19}
\nabla_{a}\,J^{abc} &=i\,\Big(
 \tfrac{8}{3} \,g^{bc} \nabla^{p}R \
{\ovp}\nabla_{p} \vp{} 
 - 16 \ \nabla^{p}\,R^{bc}\ {\ovp}  \nabla_{p}\vp
 + 8 \,\nabla^{(b}R^{c)}{}_{p}\,{\ovp} \,\nabla^{p}\vp \notag \\
&\quad \qquad -  \tfrac{4}{3}\,\nabla^{(b} R\, {\ovp} \nabla^{c)}\vp 
    + 8\ C^{pbcq} {\ovp} \nabla_{p}\nabla_{q}\vp \Big)+\text{c.c} \ , 
\end{align}
where $C^a_{\ bcd}$   is the Weyl tensor. 
An equivalent  form  of  (\ref{2.19})   found in   \cite{RT}   is   
\be
\la{2.20}
\nabla_{a}\,J^{abc} =  8\,C^{pbcq}\,\nabla_{(p} J_{q)}+  32\,\nabla_{(p}\,
C^{p bc q} \,J_{q)} \ ,    
\ee
where $J_a$ is the spin 1 current in \rf{2.10}.

The spin-3 current is  thus conserved  in  a
 conformally flat  space      but not  in a generic curved background. 
However, the important observation \ci{RT,Grigoriev:2016bzl} is that  the combined  spin 1 and spin 3  interaction term 
 \be
\la{2.21} S_{\text{int}}(h_1,h_3) = 
\int d^{4}x\,\sqrt{g}\,  (h^{a}\,J_{a}+h^{abc}\,J_{abc}) 
\ee
which is invariant under  spin 1 gauge transformation $\delta h_a = \del_a \eps$   in view of \rf{2.10} 
 can   be made invariant also under the curved-space generalization of  the spin 3  gauge transformation \rf{2.6} 
 combined   with a particular  Weyl tensor dependent    transformation  of the   spin 1 field, {\em i.e.} under 
\begin{align} 
&\delta h_{abc} =\te  \na_{(a}\,\eps_{bc)} 
-\frac{1}{3}\, g_{(ab}\,\na^{d}\,\eps_{c)d}\ , \la{222}  \\ 
&\delta h_{a} = -8\,C_{{abcd}}\,\nabla^{d}\,\eps^{bc} + 24\,\nabla^{d}C_{abcd}
\,\eps^{bc} \ . \la{2.22}
\end{align}
Note that \rf{2.20}  and \rf{2.22} simplify on an Einstein background ($R_{ab} = {1\ov 4} R g_{ab}$) as then 
$\nabla^{d}C_{abcd}=0$  and thus only one Weyl tensor  term survives.

Let us mention, as  an aside,  
 that  one may try to determine  the current  \rf{2.14}  by    imposing  $\nabla_{a}\,J^{abc}=0$   before 
 other  conditions. 
One then finds that there are no solutions    unless   one restricts   the  background to be Einstein one.
In this case 
  one finds 
 that  the coefficients in \rf{2.14}    should be 
$k_{0}=0, \  k_{2} = -\tfrac{1}{2}\,k_{1}, \  k_{4}=-\tfrac{1}{12}\,k_{1}, \ 
k_{3}=k_{5}=0. $
These values are not, however, consistent with the constraints  of Weyl invariance (\ref{2.15}) or tracelessness  (\ref{2.16}). 
Denoting    the current \rf{2.14}  with these coefficients  by 
$\widetilde J_{abc}$  we get explicitly (choosing  $k_{1}=-10$) 
\begin{align}
\la{2.24}
\widetilde J_{abc}&= -60\,i\, 
\Big[ \nabla_{(a}\,\ovp\,\nabla_{b}\nabla_{c)}\,\vp
-\tfrac{1}{2}\,g_{(ab}\,\nabla^{p}\,\ovp\,\nabla_{p}\nabla_{c)}\,\vp 
-\tfrac{1}{12}\,g_{(ab}\,R\,\ovp\,\nabla_{c)}\,\vp  
\Big]+\text{c.c}.
\end{align}
This is a non-standard current   as it does not reduce to \rf{2.2}   in the  flat space limit.
It is interesting to note that  then\footnote{Combined  with $h^aJ_a$   the coupling in  (\ref{2.25}) suggest some special role
 of the combination $h_{a}+ 6\nabla^{b}\nabla^{c}h_{abc}$; this  will   be 
discussed further  in   Appendix~\ref{app:basis}.} 
\be
\la{2.25}
h^{abc}\, \widetilde J_{abc} = h^{abc}\,\big[J_{abc}-  
6 \nabla_{(a}\nabla_{b}J_{c)}\big] \ .
\ee

\subsection{Formulation with manifest     symmetries} 

In the above discussion of the (linearized) gauge invariance   of  $S_{\rm int}$  in \rf{2.3}  or  \rf{2.7} 
we were assuming the   use of the scalar field equation, {\em i.e.} this invariance was "on-shell" one  --  valid modulo terms proportional 
to $\delta S_0 \ov \delta \vp$. 
One expects that  it should be possible  to relax  this assumption,  {\em i.e.} to extend  the invariance  to a manifest (off-shell)  one 
by  $(i)$   transforming  at the same   time  the scalar field   and $(ii)$ adding higher order terms in the  fields $h_{(s)}$. 

Let us  recall how that happens   in the simplest vector field  coupling    invariant under the  $U(1)$  gauge transformations: 
 one  introduces   the covariant derivatives  
 \be
\la{2.26}
\mathscr D_{a}\,\vp =  (\nabla_{a}+ i\,h_{a})\,\vp, \qquad\qquad 
\mathscr D_{a}\,\ovp =  (\nabla_{a}- i\,h_{a})\,\ovp\ , 
\ee
 and then the scalar   action  becomes  (here $J_0 \equiv \ovp \vp$) 
 \be \la{2.27} 
\!\!\!\!\!\! S_0 (h_1) = \int d^{4}x\,\sqrt{g}\, \ovp\,\big(\mathscr D^{2}-\six \,R\big)\,\vp
 = \int d^{4}x\,\sqrt{g}\,\Big[ \ovp\,\big(\na^{2}-\six\,R\big)\,\vp   +  h^a J_a  - h^a  h_a  J_0  \Big] \ .   
 \ee 
 This   action  which is different from the  sum $S_0 + S_{\intt}$ in \rf{2.7} 
 by an extra "nonlinear" $h^2_a$   term 
  is now manifestly  invariant under $\delta  h_a =\del_a \eps$ combined with $\delta \vp = - i \eps \vp$. 
  
 It is easy to preserve this off-shell  vector gauge invariance   in the presence of also  higher spin $s \geq 2$   
 couplings in  $S_\intt$   in   \rf{2.7}  by   just replacing $\nabla_{a}\to \mathscr D_{a}$ in the expression for the bilinear current $J^{(s)}$, thus getting 
 \be \la{2.277} 
 \sum_{s\ge 2} h^{(s)}\, J_{(s)}(\mathscr D)    =  \sum_{s\ge 2} h^{(s)} \Big[ \, J_{(s)}(\na)   +  h^a  \TT_{a (s)} +  O(h_a^2) \Big] \ . 
\ee
where $\TT_{a (s)}\equiv  (\TT_{a  b_1....b_s})$  is  a   bilinear operator   that multiplies  the term linear in the vector field   in $J_{(s)}(\mathscr D)$.

Demanding   the off-shell realization of higher  $s > 1$ spin symmetries  will require also  additional 
 non-linear terms  in the  fields $h_{(s)}$. For example, it is clear how to construct  the manifestly covariant 
  coupling  to $h_{ab}$:   one is to start   with  $S_0$ in \rf{2.7} and replace  $g_{ab} \to g_{ab} + h_{ab}$; expanding in powers of $h_{ab}$ 
  will   give  at  linear  order     the coupling to $J_{ab}$ in  \rf{2.12} (up to normalization)  plus  an infinite series of  higher order 
  terms  in $h_{ab}$. One will also be required to transform the scalar as $\delta \vp = \eps^a \del_a \vp$
   and to modify  the transformation of $h_{ab}$ in \rf{2.44}   by  order $h_{ab}$ terms   to recover the usual   form of transformation of $g_{ab} + h_{ab}$  under the diffeomorphisms.
   
   Similarly, for spin 3   one will need to  supplement  the transformations   in \rf{222},\rf{2.22}  
   with a transformation of  the scalar  field  to cancel the terms proportional to $\delta S_0 \ov \delta \vp$ that were dropped in \rf{2.20}; that will then require adding also $(h_{abc})^2$ terms in the  action $S_\intt$ 
   to compensate for the variation of the $h_{abc} J^{abc}$ term under this transformation of $\vp$, 
   {\em etc.} 
   
   An  alternative to this  procedure is  to follow  the approach of 
   \ci{Segal:2002gd}  (in flat   case)   and  \ci{Grigoriev:2016bzl} (in curved background) 
   and introduce only linear $h^{(s)}\, J_{(s)}$  couplings  but to the
    whole tower of the   higher   spin fields including the scalar $h_0$ coupled to $J_0 = \ovp \vp$  and transform  both $\vp$  and  $h_{(s)}$. 
    In this case the   gauge transformation   of  $h_{(s)}$  will  contain,  in addition to $\nabla \eps_{(s-1)}$  term, 
     also   terms  linear in $h_{(s')}$  \ci{Grigoriev:2016bzl}. 
   The two approaches  should   be  related by  field redefinitions like $ h_0 \to h_0 -  h^a h_a$
   and so on,   ({\em cf.}   \ci{Beccaria:2016syk}). 
         
Starting  with an action $S(\vp, h)$  which contains all necessary  
terms to be manifestly 
invariant under some  local  transformation\foot{Here   
 $\eps$  may stand for 
parameters of Weyl, gauge, or algebraic symmetries as in \rf{2.9},\rf{2.44}.}
$\delta \vp = F(\eps; \vp, h),\ \  \delta h = f(\eps; h)
$, 
and then   integrating out $\vp$   one should    get 
the  (full, non-local) effective action  $\Gamma$ in 
\be\la{2.28} 
Z= e^{-\Gamma(h)} = \int d\vp\,e^{-S(\vp, h)} \ ,
\ee
which should be formally  invariant  
under $\delta h = f(\eps, h)$. 
As $\G$  is given just  by a 1-loop determinant  ($\vp$  does  not have self-interactions)  
its  logarithmically  UV singular part    is  local  and  
 cannot  contain any anomalies
\be \la{2.29} 
\Gamma(h) = \log \Lambda_{_{\rm UV}}\, 
\SS(h)+\dots \ . 
\ee
Thus   $\SS(h)$  (that we shall call the {\em induced} action) 
  should   be  
manifestly invariant  under  the above transformations of $h$.

Suppose  we start instead with an action $\wt S(\vp, h)= S_0 (\vp) + h \cdot J $ that  contains only linear  in $h$ terms and is invariant under $\delta h = f(\eps; h)$ only on-shell, {\em i.e.} up to terms proportional to  the free $\vp$ equation of motion. As 
the terms proportional to the equations  of motion  contribute delta-functions  to the coordinate-space 
correlators  
of $J$,   they   can be ignored  as usual  in  the correlation functions
at {separated} points  which will thus   be  invariant. 
 However, the corresponding  local induced action  $\wt \SS(h)$ 
   is no longer guaranteed  to be invariant   under $\delta h = f(\eps; h)$.

Indeed, in the vector coupling case ({\em cf.} \rf{2.27}) 
  it is easy to see  
 that starting just with the minimal 
$h^a J_a$ coupling term  one gets the induced  action containing  non-invariant $(h^{a}h_{a})^{2}$ term.
Same will   happen for higher spin  couplings. It should be possible to 
eliminate 
such non-invariant   terms  by  a field redefinition  provided 
one  considers   $\wt \SS(h)$  for  the whole tower of the conformal  higher spin fields.
For example,  including 
 non-zero   scalar $h_0  $  we will get  the term $(h_0 + h^a h_a)^2$   and thus  non-invariant 
$(h^{a}h_{a})^{2}$  term  can be redefined away   by a shift of $h_0$.
This has, of course,  an explanation  in terms of the off-shell  invariance of the action $S(\vp,h)$ 
 in \rf{2.27} that has  the term $h^a h_a J_0$ present  there. 
Similar  observations   should apply to higher spin cases as well. 


\section{Structure of the  induced  action} 
\la{sec:ind}

Starting  with the reparametrization and  vector gauge invariant  conformal  scalar action \rf{2.27}  and integrating $\vp$ out  the 
resulting  induced action for $g_{ab}$ and $h_a$ ({\em i.e.}  the  coefficient of the logarithmic UV divergence in the effective action \rf{2.29}) 
will  take the familiar form\footnote{This  expression is given by the  relevant 
Seeley-de Witt   coefficient  
of  the conformally coupled scalar   Laplacian, see, 
{\em e.g.},  \cite{Hooft:1974bx}.  Here  we dropped  a total derivative 
$\sim R^{\star}\,R^{\star}$ term.}  
\be
\la{3.3}
\text{S} = \int d^{4}x\,\sqrt{g}\,\Big(
\te  -\frac{1}{12}\,F^{2}_{ab}  
+\frac{1}{120}\,C^{2}_{abcd}\Big) \ , 
\ee
where $F_{ab}=\del_{a} h_{b}-\del_{b} h_{a}$  and $C_{abcd}$ is the Weyl tensor. 
$\SS$  is  invariant under the reparametrizations, vector gauge  symmetry and the Weyl symmetry. 

One  may systematically obtain $\SS$  by expanding  $\exp(-S)$ in \rf{2.28}  in powers of the 
fields $h_{(s)}$    and computing the UV    singular parts of the resulting 
 correlators  of the currents   on a curved   background using, {\em e.g.},  
the  covariant methods of \cite{Christensen:1976vb,Brown:1977pq,Barvinsky:1985an,Barvinsky:1994cg,Jack:1983sk,Avramidi:1986mj,Osborn:1989bu,Avramidi:1990je,Herman:1995hm,Aida:1996zn}.
For example,   computing the correlator of the vector current $\langle J^{a} J^{b}\rangle$ 
 using the  dimensional regularization expressions   in Appendix  A of \cite{Jack:1983sk}
one finds 
\be
\la{3.5}
\,\int d^{4}x \, \sqrt{g}\, h^{a}\,\langle J_{a}\,J_{b}\rangle_{_{\text{UV}}}\,h^{b} = 
-{\te \frac{1}{6}}\,\int d^{4}x\,\sqrt{g}\ F^{2}_{ab} \ . 
\ee
Here and below $\langle ...\rangle_{_{\text{UV}}}$   will stand for the   coefficient of the logarithmically divergent 
(or  pole   ${1/ \varepsilon} \sim \log \Lambda_{\rm UV}$) part of  a  correlator.
The $h^a h_a$ term in the manifestly  gauge-invariant action \rf{2.27} does not  contribute at  $h^2$ order
as $\langle J_0 \rangle_{_{\text{UV}}}=0$; it produces  the $h^4$ term   that cancels, however, against 
other  $h^4$   contributions so   the final result is in agreement with \rf{3.3}. 

A similar approach can be used in the spin 2 case.   
  If one  starts with  the  scalar action $S_0$ in \rf{2.7} and adds   just a linear  coupling term $h^{ab} J_{ab}$   then 
 the coefficient of the linear in $h_{ab}$ term in  the corresponding induced action 
 $\SS$   will   be   given   by  the UV log divergent part of the 1-point function of the  
  spin 2 current, {\em i.e.} 
  $\langle J_{ab} \rangle_{_{\text{UV}}}  $, which turns out  ({\em cf.} \rf{4.9},\rf{4.11}) to be   proportional to the   Bach tensor $ B_{ab}$
  (defined  in  Appendix~\ref{app:bach}). 
  To reproduce the  gauge-covariant  quadratic   $  h_{(2)}  \mc O_4 h_{(2)} $   term  one  will need,   in general,   to add   to 
   $\langle J_{ab}  J_{cd} \rangle_{_{\text{UV}}} $  also the 
     "contact term"  contribution of the  quadratic   $h^{ab}  \TT_{abcd}  h^{cd}$  coupling 
    term in the  manifestly gauge-invariant  scalar action).\foot{Here  $\TT_{(4)}$ is a scalar bilinear  operator  containing 2 derivatives.
      As discussed   above, 
   similar non-linear  in $h_{(s)}$ terms in the scalar  action $S(\vp, h)$  can be reconstructed 
    order by order by demanding its manifest (off-shell)   invariance.} 

  In the spin 2 case there is a short-cut: 
   we   may  shift the metric $g_{ab} \to g_{ab} + h_{ab}$  in  the conformally coupled scalar   action  to 
 isolate the  spin 2  coupling as in \rf{2.7},\rf{2.11}; then 
   the resulting dependence  of $\SS$ on $h_{ab}$    should  be given by the expansion of the   $C^2$ term in \rf{3.3}
 \be \la{3.1} 
 \int d^4 x \sqrt g \,  C^2_{abcd}  \to   \int d^4 x   \sqrt g \Big[  B_{ab}(g)  h^{ab}  +   h^{ab} \OO_{abcd} (g) h^{cd} + ...\Big]  \ ,  \ee 
 where   $B_{ab}$ is    the Bach tensor   and  the operator $\OO_4 \sim \na^4 + ...$ is 
 reparametrization and Weyl invariant and  invariant under the algebraic symmetry $\delta h_{ab} =  g_{ab} \xi$.
 The invariance of the $\OO_4$ term under the gauge symmetry $\delta h_{ab} = \na_{(a} \eps_{b)}$   requires 
  the cancellation of the linear in $h_{ab}$ term in \rf{3.1}, {\em i.e.}   constraining  $g_{ab}$ 
    by  the condition of Bach-flatness,  $B_{ab}=0$. 
We also   note that  the   quadratic coupling term $h^{ab} \TT_{abcd} h^{cd}  $ in the covariant scalar action   found   by   shifting 
$g_{ab} \to g_{ab} + h_{ab}$  in $S_0$ in \rf{2.7}  and expanding  in $h_{ab}$  will contain 
$(i)$ part   with derivatives acting on 
$h$ (coming from $R/6$-term)  and thus giving zero contribution  as  $\langle J_0 \rangle_{_{\text{UV}}}=0$, and $(ii)$  part 
with two derivatives   acting on the scalar field   and thus similar  to \rf{2.11}, with  the resulting contribution again proportional to $B_{ab}$. 
Thus  its contribution can be ignored  on the Bach-flat backgrounds.

 Similar   remarks apply  to  higher spin coupling terms. 
In general,  the expansion of  the induced action $\SS(g,h)$ in  powers of  $h_{(s)}$
should be \cite{Grigoriev:2016bzl}
\begin{align}
\label{310}
 & \SS(g,h)=\SS^{(0)} (g)+\SS^{(1)}  (g,h)+\SS^{(2)}  (g,h)+\ldots\,,\qquad \\
 & \SS^{(1)}=\sum_s   \int d^4 x \sqrt{g} \,B_{(s)}(g)\ h^{(s)}\,, \qquad  \qquad
  \SS^{(2)} =\sum_{s,s'}  \int \  h^{(s)} \ \OO_{s,s'} (g) \ h^{(s')} 
  \ , \ \ ...
  \la{3100}
 \end{align}
 This action should have manifest  reparametrization and Weyl  symmetries. 
 $\SS^{(0)} (g)$ is the Weyl  tensor  term in \rf{3.3} (while spin 1 term in \rf{3.3} is included in $\SS^{(2)}$). 
 
Ignoring  total derivatives,    the coefficient   of the   linear term
 $\langle J_{(s)  }  \rangle_{_{\text{UV}}} \sim 
 B_{(s) }(g)$    should   be 
a local function of the metric $g$  and its derivatives, which is covariantly conserved, traceless and Weyl-covariant. 
Explicitly, $B_a=0$, $B_{ab}$ is the    Bach tensor   and as we will show in Appendix \ref{app:linear}
$B_{abc}=0$  for any background.  

As was argued  in \cite{Grigoriev:2016bzl},   for general $s$, the tensor $B_{(s) }(g)$   should vanish  on a Bach-flat background, at least 
up to terms quadratic in the curvature of the background metric. 
The vanishing  of $B_{(s)}(g)$   is required in order for the Bach-flat  metric $g_{ab}$ 
along  with $h_{(s)}=0$   be the vacuum of  the full CHS action $\SS(g,h)$.
In that case the quadratic term  $\SS^{(2)}$   which, in general,  is non-diagonal   in $s,s'$,  
 should be   invariant under the background-covariant  gauge and algebraic   transformations 
of the CHS fields generalizing \rf{2.44} (like  \rf{2.2}, {\em etc.}). 

The  operator $\OO_{s,s'} (g)$  should, in general, 
 receive contribution from $\langle J_{(s)  } J_{(s')  } \rangle_{_{\text{UV}}}$
as well as  from the   contact term   $ X_{(s) (s')  }  =  \langle \TT_{(s) (s')  } \rangle_{_{\text{UV}}}$    
 coming from  the quadratic term   $h^{(s)} \TT_{(s)(s')} h^{(s')}  $ required for  the manifest  covariance of  the 
  scalar action  ({\em cf.} \rf{2.27},\rf{2.277}). 
  
  We shall   make  the  conjecture that  $ X_{(s) (s')  }=0$ on a Bach-flat background. 
  As was mentioned above,   this is true   in spin 2 case  where    $ X_{(2) (2')  }$  is proportional to the Bach tensor. 
  In general,  since the dimension of  the CHS  field  $h_{(s)} $ is $2-s$  (so that  the interaction action 
  in \rf{2.3} is dimensionless) 
   the product  $h_{(s)} h_{(s')}$     has   the  same  dimension  as $h_{(s'')}$ with 
   $s''= s+s' - 2$, 
   and thus it   may  be  possible   to eliminate   the $h^{(s)} \TT_{(s)(s')} h^{(s')}  $ term  
   by a redefinition of 
      $h_{(s'')}$ in the $h^{(s'')}J_{(s'')}$ coupling. As the linear term   in $h^{(s'')}$ in $\SS_1$ in \rf{3100}     should vanish on a 
      Bach-flat   background,   the same  should then apply to the  contribution  of 
      $h^{(s)} \TT_{(s)(s')} h^{(s')}  $ 
      to  $\OO_{s,s'} (g)$. 
        Equivalently, as 
    the scalar bilinear operator $X_{(s) (s')  }$ has the same dimension as $ J_{(s'')}$ it is natural  
    to expect that the (reparametrization and  Weyl  covariant)  expectation value 
      $ X_{(s) (s')  }  =  \langle \TT_{(s) (s')  } \rangle_{_{\text{UV}}}$  should also vanish 
      on a Bach-flat background, as it happened in the case of 
         $\langle J_{(s'')}  \rangle_{_{\text{UV}}}$.
For example,  like the 
$h_a^2 $  term  in  \rf{2.27} can be  absorbed into a redefinition of the CHS scalar $h_0$, 
a  possible  $h_{(1)} h_{(3)}$ term  in  \rf{2.277}   may  be absorbed into   a redefinition of $h_{(2)}$. 
Indeed, we will  check   below  that  $ X_{(1) (3)  }  =  \langle \TT_{(1) (3)  } \rangle_{_{\text{UV}}}$ 
vanishes on a  Bach-flat     background. 
Similarly, it should be possible to absorb the $ h^{(3)}\TT_{(3)(3)} h^{(3)}$ term in the scalar action  
into $h^{(4)}J_{(4)} $    
 so that  its tadpole contribution    
   should be proportional to the variation of a linear term $B_{(4)}(g)\,h^{(4)}$  and  
   should  thus vanish along with $B_{(4)}$ if $B_{ab}=0$.

\section{Spin 3 induced action}
\la{sec:ind-3}

Below   we will study in detail the dependence of the quadratic part  $\SS^{(2)}$ of the  induced action on the spin 3 field, and, in particular, its mixing   with the spin   1  field   anticipated in  \cite{Grigoriev:2016bzl}.
Our starting  point will be the manifestly  vector gauge  covariant form of the scalar action \rf{2.27},\rf{2.277}. 
We will   choose  $h_{abc}$  to be traceless.

As  the linear in $h_{(3)}$ term in the  induced action in \rf{3100} vanishes  
(as shown in    Appendix~\ref{app:linear}, $\langle   J_{abc} \rangle_{_{\rm UV}}$=0)
the  induced action  in the  spin 1  plus  spin 3  sector starts  with a quadratic term   
\be
\la{4.1}
\text{S}^{(2)}  = \text{S}_{11}+\text{S}_{13}+\text{S}_{33} \ , 
\ee
where 
 $\text{S}_{ss'}$ is a term bilinear in $h_{(s)}$ and $h_{(s')}$. 
$\text{S}_{11}$ is the Maxwell action in  \rf{3.3},(\ref{3.5}). 
The 1--3  mixing term will have  two contributions:
\be \la{411}
\text{S}_{13}=\text{S}_{13}^{\text{(a)}}+\text{S}_{13}^{\rm (b)}\ , \ee 
  where 
$\text{S}_{13}^{\text{(a)}}$  will come from the correlator $\langle J_{(1)}\,J_{(3)}\rangle_{_{\text{UV}}}$  and 
 $\text{S}_{13}^{\rm (b)}$   from the contact term $ \langle \TT_{(1)(3)}\rangle_{_{\text{UV}}}     $   (see \rf{2.277}). 
Similarly, the $\text{S}_{33}$ term
\be \la{333}
\SS_{33} = \int d^4 x \sqrt g \  h_{(3)} \mc O_6\,  h_{(3)}\ , \ee
  will  come   from  the correlator $\langle J_{(3)}\,J_{(3)}\rangle_{_{\text{UV}}}$ 
and also from the 
 contact term $ X_{(3) (3)  }  =  \langle \TT_{(3) (3)  } \rangle_{_{\text{UV}}}$  
 that   originates  from the $h_{(3) } \TT_{(3) (3)  }  h_{(3) } $   term in the manifestly covariant scalar action.


\subsection{Spin 1--3  mixing   term} 

A long straightforward calculation  using covariant methods  of \cite{Christensen:1976vb,Brown:1977pq,Barvinsky:1985an,Barvinsky:1994cg,Jack:1983sk,Avramidi:1986mj,Osborn:1989bu,Avramidi:1990je,Herman:1995hm,Aida:1996zn}
shows that the  contribution to $\text{S}_{13}$   coming from the UV singular 
 part of the  2-current correlator  is  given by  (see Appendix \ref{ab} 
for some details of this computation) 
\be
\la{4.2}
\text{S}_{13}^{\rm (a)}=
\int d^{4}x \, \sqrt{g}\, h^{a}\,\langle J_{a}\,J_{bcd}\rangle_{_{\text{UV}}}\,h^{bcd} = 
\int d^{4}x\,\sqrt{g}\ \mathscr L_{13}^{\rm (a)}\ ,
\ee 
where 
\begin{align}
\la{4.3}
\!\!\! \mathscr L_{13}^{\rm (a)} = &
- \tfrac{74}{15} R^{bc} R h^{a} h_{abc} -  \tfrac{28}{15} h^{a} \
C_{b}{}^{def} C_{cdef} h_{a}{}^{bc} + 
\tfrac{116}{15} h^{a} C_{b}{}^{def} C_{cedf} \
h_{a}{}^{bc} \nonumber \\ 
& + \tfrac{74}{5} R_{b}{}^{d} R^{bc} h^{a} h_{acd} -  
\tfrac{38}{5} R^{bc} h^{a} C_{bdce} h_{a}{}^{de} + 
\tfrac{4}{5} R^{bc} h^{a} C_{adce} h_{b}{}^{de} \nonumber \
\\ 
& - 8 h^{a} C_{a}{}^{e}{}_{b}{}^{f} C_{cedf} \
h^{bcd} + \tfrac{1}{6} h^{a} \nabla^{b}R \nabla_{c}h_{ab}{}^{c} \
+ \tfrac{13}{3} h^{a} h_{abc} \nabla^{c}\nabla^{b}R \nonumber \\ 
& + 11 h^{a} \nabla_{a}R^{bc} \nabla_{d}h_{bc}{}^{d} - 11 h^{a} 
\nabla^{c}R_{a}{}^{b} \nabla_{d}h_{bc}{}^{d} -  \tfrac{37}{5} h^{a} \
h_{abc} \nabla_{d}\nabla^{d}R^{bc} + 10 h^{a} \nabla_{c}h_{abd} \
\nabla^{d}R^{bc} \nonumber \\ 
& - 10 h^{a} \nabla_{d}h_{abc} \nabla^{d}R^{bc} + \tfrac{56}{5} \
h^{a} h_{bcd} \nabla^{d}\nabla_{a}R^{bc} -  \tfrac{56}{5} h^{a} \
h_{bcd} \nabla^{d}\nabla^{c}R_{a}{}^{b} \nonumber \\ 
& + 6 h^{a} \nabla_{b}h^{bcd} \nabla_{e}C_{acd}{}^{e} + \
12 h^{a} \nabla^{d}h_{a}{}^{bc} \nabla_{e}C_{bdc}{}^{e} \
+ \tfrac{32}{5} h^{a} h^{bcd} \nabla_{e}
\nabla_{d}C_{abc}{}^{e} \nonumber \\ 
& + \tfrac{24}{5} h^{a} h_{a}{}^{bc} \nabla_{e}
\nabla_{d}C_{b}{}^{d}{}_{c}{}^{e} + 8 h^{a} 
\nabla_{d}C_{abce} \nabla^{e}h^{bcd} + 8 h^{a} \
C_{acde} \nabla^{e}\nabla_{b}h^{bcd} \nonumber \\ 
& + 8 h^{a} C_{bdce} \nabla^{e}\nabla^{d}h_{a}{}^{bc} \ . 
\end{align}
The action  \rf{4.2}  is  invariant under the Weyl transformations \rf{2.9} 
but is not  vector gauge  invariant: to restore this  invariance   we need to add the "tadpole"  contribution of the 
mixed 1-3 term in \rf{2.27},  {\em i.e.}  
\begin{align}
\la{4.5}& \qquad \qquad \qquad   \mathscr L_{13}^{\rm (b)} = \langle \mc T \rangle_{_{\text{UV}}} \ , \ \ \ \ \ \ \ \ \ \ \ 
\\
\la{4.7}
\mc T  \equiv & h^{d}\,\TT_{dabc}\,h^{abc} = \big( - 12\, h^{a}  \nabla_{c}
\nabla_{b}h_{a}{}^{bc} +  84\, R^{bc}\,h^{a}  h_{abc} \big) \ovp \vp\, - 60\,h^{a}\,\,\nabla^{c} h_{abc} \  \nabla^{b} (  \ovp \vp 
 ) \notag \\
&  \ \ \ \qquad \qquad \qquad 
 + 120\, h^{a} h_{abc}\big[  \nabla^{b}\ovp \nabla^{c}\vp  -  ( \ovp \nabla^{b}\nabla^{c}\vp +  
 \vp \nabla^{b}\nabla^{c}\ovp ) \big] \ .
\end{align}
To  compute  the tadpole contribution\foot{In a massless theory, one  has to be careful to 
isolate the   UV poles from the IR ones. 
In  curved space case, the  curvature plays  the role of an effective  IR scale 
(which can be   captured by a  resummation of an infinite set of terms in near-flat-space expansion). 
Taking this   into  account, the  tadpoles   lead to non-trivial contributions to  logarithmic UV divergences.
Various point-splitting treatments of tadpoles in curved space are discussed in 
\cite{Christensen:1976vb,Adler:1976jx,Adler:1977ac,Wald:1978pj}.}
 $\langle \mc T \rangle_{_{\text{UV}}} $
  we note that for a conformally coupled scalar 
 the  pole part of the 2-point function vanishes, 
$\langle \ovp\vp\rangle_{_{\text{UV}}}=0$ (see, {\em e.g.},   \cite{Wald:1978pj}).
Then dropping a total derivative  term  we get 
\be
\la{4.9}
\langle \mc T \rangle_{_{\text{UV}}} =  60\,h^{a}\,h_{abc}\,\langle J_{ab}\rangle_{_{\text{UV}}} \ , 
\ee
where $J_{ab}$ is the   stress tensor of the conformal scalar defined  in \rf{2.12}. 
The  UV singular part of the expectation  value of the  scalar stress tensor   should be related to 
the   derivative of the $C^2_{abcd} $  logarithmic  divergence  in the effective action ({\em cf.} 
\rf{3.3}) and should thus
 be proportional to the Bach tensor in \rf{A.2}.   Indeed, as follows  from 
  eq.(6.4) in  \cite{Christensen:1976vb},\foot{See also \cite{gibbons1979quantum} for a detailed discussion 
of regularization issues in the vacuum expectation value of the stress tensor.} 
\begin{align}
\notag
\mathscr L_{13}^{\rm (b)} =  \langle \mc T \rangle_{_{\text{UV}}} = & 6\,B^{ab} h_{abc}\,h^{c}\,\\
\la{4.11}
= &  \big(    -2\, R^{ab} R + 6\,  R^a_d R^{bd}  + 6 R_{ed}   C^{e ab d}  + \nabla^{a}\nabla^{b}R   - 3 \nabla^2 R^{ab} 
\big) h_{abc}\,h^{c} \ . 
\end{align}
Thus   the tadpole contribution vanishes on a Bach-flat background   in agreement  with the above discussion.

Adding \rf{4.11} to (\ref{4.3}) we  get  a simple  manifestly  vector   gauge  invariant  expression 
\begin{align}
 \mathscr L_{13} = \mathscr{L}_{13}^{\rm (a)}+\mathscr L_{13}^{\rm (b)} = 
8\, F^{ab}\,\Big[C_{a}{}^{cdp}  \, \nabla_{p}h_{bcd} + \big( \nabla_{a}R^{cd}   
-\, \nabla^{c}R_{a}^{d}\big)   h_{bcd} \   
 \Big]\ ,  \la{4.12}
\end{align}
where $F_{ab} = \del_a h_b - \del_b h_a$  and $h_{abc}$ is  totally symmetric and  traceless. 
 Like  each of the  $ \sqrt g \mathscr{L}_{13}^{\rm (a)}$ and $\sqrt g \mathscr L_{13}^{\rm (b)} $ 
 terms, 
 their sum  $\sqrt g  \mathscr L_{13} $ 
  is also 
  invariant under the Weyl transformations ({\em cf.} also the discussion at the end of 
  Appendix~\ref{app:linear}).
  Both $\mathscr{L}_{13}^{\rm (a)}$     and $\mathscr{L}_{13}^{\rm (b)}$  vanish on a
   conformally-flat Einstein space.

  \subsection{Spin 3   gauge invariance}
\la{sec:g-ein}

As we  saw in section 2, the   spin 3 interaction term \rf{2.13}  is not  invariant under the 
curved background  spin 3 gauge transformation   \rf{2.44}   --  we need also to transform 
the spin 1  field in the 
 interaction term in \rf{2.7}  according to \rf{2.22}.
 If  we  specify the metric to be, {\em e.g.}, 
    the Einstein  one  ({\em i.e.} a particular  Bach-flat  one)   then   $\nabla^{a}C_{abcd}=0$
 and \rf{222},\rf{2.22} simplify  to 
\be
\la{4.16}  \delta h_{abc} =\te  \na_{(a}\,\eps_{bc)} 
-\frac{1}{3}\, g_{(ab}\,\na^{p}\,\eps_{c)p}\ , \qquad \qquad 
\delta h_{a} = - 8 \,C_{abcp}\,\nabla^{p}\,\eps^{bc}  \ .
\ee
Using \rf{3.3} and that $R_{ab}= {1\ov 4 } R g_{ab} $ in \rf{4.12},  the  quadratic part of the induced 
action \rf{4.1}   may then be written as 
\be
\la{4.17}
\text{S}^{(2)} 
 = \int d^{4}x\,\sqrt{g}\,\Big[\te -\frac{1}{12}\,F^{2}_{ab}   + \ 8\, C^{abcd}\,F_{ap}\,\nabla_{d}h^p_{\, \ bc} \ 
+ \ h_{(3)}\,\mc O_6 \, h_{(3)}\Big] \ . 
\ee
One can then   check that the   transformation of  $h_a$  in \rf{4.16}  in the first term in \rf{4.17}   
 combined with  the transformation  of $h_{abc}$ in the   second  mixed term  in \rf{4.17}  leaves the total action 
invariant (see Appendix~\ref{app:check-gauge}).

 The  transformation of  the     second term in \rf{4.17} under the  transformation 
 of $h_a$  should cancel against  the  transformation of the last  pure spin 3 term in \rf{4.17} under the variation of $h_{abc}$ in \rf{4.16}.  
 As a  result,  the  last term   $\SS_{33}$ in  the action \rf{4.1}   (and thus the operator $\OO_6\sim \nabla^6 + ... $ in \rf{4.17})
     cannot be,   in general,    invariant under 
 the spin 3 gauge transformations  even on an Einstein background, contrary to what one might naively 
 expect.\footnote{See Appendix~\ref{app:basis} for a  discussion of how  this  may    
 change if one  requires  only  the  invariance under the restricted 
spin 3 gauge transformations with $\nabla_{a}\eps^{ab}=0$.} 

Since the variation of $h_a$ in \rf{4.16}  is linear  in the Weyl tensor,    the variation of the second  term in \rf{4.17}   
has the structure $ C \nabla ( C \nabla \eps) \nabla h_{(3)}$, {\em i.e.}  is  of {\em second} order in the curvature. 
Thus  the $h_{(3)}\,\mc O_6 \, h_{(3)}$    term may  be invariant on its own  at leading  linear order in the curvature
in a  small curvature expansion. 
Indeed, such an  operator   was  constructed  in \ci{Nutma:2014pua}
   starting  from the condition of such linearized spin 3 gauge invariance.
    Below   we shall reproduce  this  result  by directly  computing  the  leading term in  the induced action 
    $\SS_{33}$  
     in  the near-flat-space expansion.\foot{The  operator $\OO_6$ found in  \ci{Nutma:2014pua}
     was not unique, so the matching to our  result for the  induced action 
     requires fixing   the remaining freedom  in  \ci{Nutma:2014pua} in a particular way.}



\subsection{Pure spin 3 term} 
\la{sec:NT}

As was discussed  above, the  term $\text{S}_{33}$ in \rf{4.1}  may receive contributions from 
$(i)$ the correlator $\langle J_{(3)}(x)\,J_{(3)}(x')\rangle_{_{\text{UV}}} $ 
and  $(ii)$  the  tadpole   term $ X_{(3) (3)  }  =  \langle \TT_{(3) (3)  } (x) \rangle_{_{\text{UV}}}$    coming from 
the  quadratic  $h_{(3) } \TT_{(3) (3)  }  h_{(3) } $   term in the manifestly  spin 3   gauge-covariant scalar action. 
The latter should  vanish on a Bach-flat background   as discussed above.

The exact  computation  of  $ \langle J_{(3)}(x)\,J_{(3)}(x')\rangle_{_{\text{UV}}} $    with  spin 3 current given in \rf{2.18}
in the background-covariant   approach  is 
technically   challenging  and will not be attempted here. We shall discuss   only the flat space case and 
the near-flat-space expansion to leading order in the curvature 
making contact with an  earlier result of \ci{Nutma:2014pua}.

 In the flat space limit, we have
\be
\la{4.13}
\text{S}_{33}^{\rm flat} = \int d^{4}x\, \mathscr L_{33}^{\rm flat}  \ ,   \qquad \qquad 
 \mathscr L_{33}^{\rm flat}=   \ha \,h^{abc}\,
\langle J_{abc}\, J_{a'b'c'}\rangle_{_{\text{UV}}}^{\rm flat} \  h^{a'b'c'}.
\ee
The    correlator  $\langle J_{(3)}(x)\,J_{(3)}(x')\rangle^{\rm flat}$ 
of the  flat-space  spin 3 current  \rf{2.2}    computed in the  free  scalar CFT is given by the   transverse traceless  spin 3 projector  operator
times  $|x-x'|^{-6}$. To extract the UV pole we may use, {\em e.g.},  the dimensional
regularization  as in  \cite{Jack:1982hf}. The resulting expression
 is\footnote{The computation amounts to the evaluation of
$\partial_{a_{1}}\partial_{a_{2}}\cdots G(x-x')\,
\partial_{b_{1}}\partial_{b_{2}}\cdots G(x-x')$,
where  $G$ is the free scalar propagator
$G(x) = \frac{1}{8\,\pi^{2}\,\sigma}$ with $\sigma = \frac{1}{2}\,(x-x')^{2}$.
Taking   derivatives and using  the relation
$\partial^{2}{\sigma^{-p}} = {2\,p\,(p-1)}{\sigma^{-p-1}}$,
we may  reduce all terms in (\ref{4.13}) to the form
$h^{abc}(x)\,h^{a'b'c'}(x')\, P_{abc,a'b'c'}(x-x')\,
(\partial^{2}_{x})^{k}\,\frac{1}{\sigma^{2}}$
where $P(x-x')$ is a tensor built with the  displacement vector $(x-x')_{a}$, and
$\frac{1}{\sigma^{2}}\sim \frac{1}{\eps}\,\delta^{(4)}(x-x')$ gives the  pole in dimensional   regularization 
(see,  for instance,  eq.~(A.1) of \cite{Jack:1983sk}).
The final result is obtained by integrating by parts the
 $\del_{x}^2$ operators and taking in the end
the coincidence limit $x\to x'$.}
\begin{align}
\la{4.14}
\!\!\!\!\!\!\mathscr L_{33}^{\rm flat}   =\te  \frac{7}{45 } \big(h^{abc} \Box^{3}\,h_{abc}   -\tfrac{2}{5}\, h^{abc}\, \partial_{abcdef}\,h^{def} 
+ \tfrac{12}{5}\, h^{abc} \Box\,
\partial_{bcde}\,h_{a}{}^{de}  
 -  3\,h^{abc}\, \Box^{2}\,
\partial_{cd}\,h_{ab}{}^{d}    \big) \ .
\end{align}
As expected, the   spin 3 CHS Lagrangian (\ref{4.14}) is  invariant under the gauge transformations \rf{2.4}.\footnote{Notice
that the gauge invariance fixes the coefficients in (\ref{4.14}) up to an overall proportionality constant. }
It is  also   scale-invariant, and,  being  a flat limit of a  full Weyl-invariant  action, 
it should also   have  the full conformal  symmetry. 
 This is  indeed  the case as    we demonstrate in 
  Appendix~\ref{app:T3}:   the Lagrangian (\ref{4.14})
admits  a symmetric traceless stress tensor which  is conserved and gauge invariant  on the   spin 3 equations of motion. 

Next, we may compute  the first correction to \rf{4.14} in the near-flat-space expansion, {\em i.e.}  at the leading order in 
$h_{ab}= g_{ab} - \delta_{ab}$. Schematically,
\begin{align}
\la{4.19}
\text{S}_{33} = \text{S}_{33}^{\rm flat}+\text{S}_{233} +\cdots, \qquad \quad 
\text{S}_{233} = \int d^{4}x\,h_{(2)}\,\big[h_{(3)}\,\partial^{6} h_{(3)}+
 \partial h_{(3)}\,\partial_{5}\,h_{(3)}+\cdots\big] \ .
\end{align}
Since the transformation of $h_a$ under the spin 3 gauge transformations  given in \rf{2.22}  
involves already one  power of the  curvature,  the term $\text{S}_{233}$   can not mix with the spin 1-3  term $\SS_{13}$ 
and should thus be   invariant under the linearized spin 3 gauge transformations on its own. 

\begin{figure}[t]  
  \centering
  \ifLocalFigs
   \includegraphics[scale=0.17]{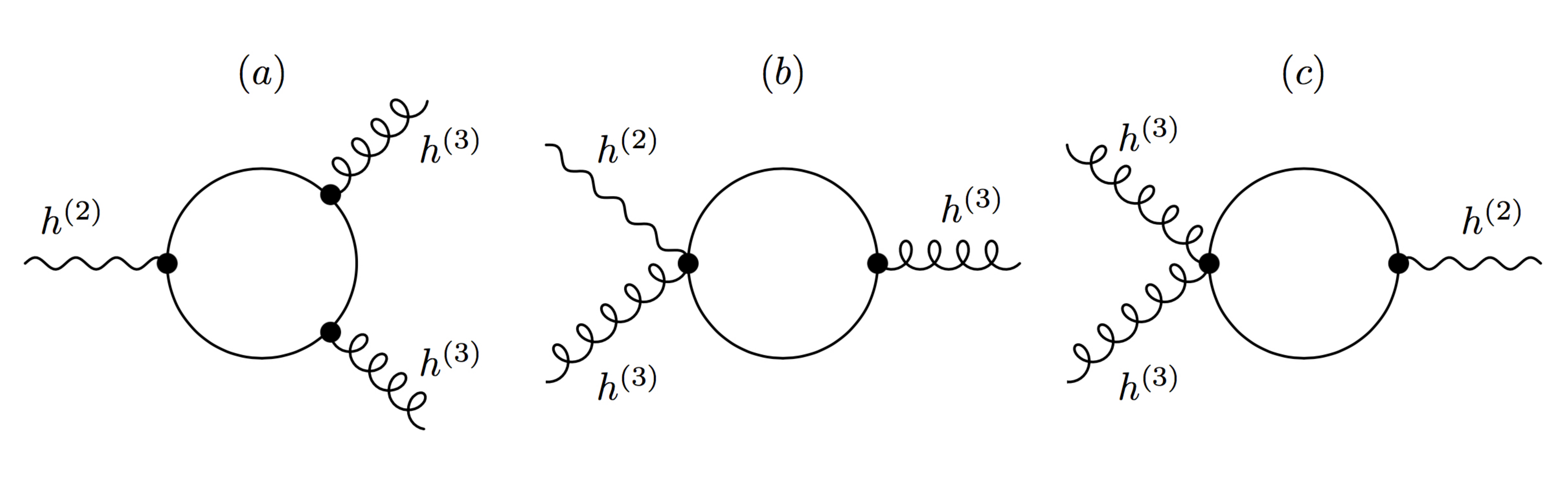}
  \else
 \includegraphics[scale=0.17]{./Figures/FeynArts/233-abc.pdf}
 \fi
   \caption[]
   {\label{fig:233} Scalar field one-loop   diagrams  with  UV divergent parts contributing  to 
    the induced action 
   $\text{S}_{233}$. Dots stand for the insertions  of the spin 2 or spin 3  flat-space  currents in \rf{22},\rf{2.2}  or 
   the quadratic $h_{(2)} h_{(3)}$   and $h^2_{(3)}$  vertices  in the scalar   action  required for  its  manifest   covariance.
   }
\end{figure}

The first correction  $\text{S}_{233} $   is given by the sum of the 
three  contributions shown in Fig.~\ref{fig:233}, {\em i.e.} 
\be
\la{4.20}
\text{S}_{233} = \text{S}_{233}^{(a)}+ \text{S}_{233}^{(b)}+ \text{S}_{233}^{(c)}.
\ee
To simplify the computation we shall assume that the external fields $h_{(2)}$ and $h_{(3)}$ are transverse and traceless (TT).  
  The first  diagram (a) has  three flat-space  current  vertices
that have a  simple form $h^{(s)}\,\ovp\,\partial^{s}\,\vp$  
(see \rf{22},\rf{2.2}  and  \cite{Beccaria:2016syk}).
The explicit form of the 
 vertex $h_{(2)}  h_{(3)} \ovp \vp $ in the  diagram (b)
  is found  by expanding the curved-space source term (\ref{2.13}) in powers of $h_{ab}$
\be
\la{4.21}
\int d^{4}x\,\sqrt{g}\,h^{abc}\,J_{abc}(\nabla) = \int d^{4}x\,h^{abc}\,J_{abc}(\partial) +
\int d^{4}x\, h^{ab}\mc V_{ab} (h_{(3)}, \vp) +\mc O(h_{ab}^{2})\ , 
\ee
where  the 
 explicit form of $\mc V_{ab}$ is  given in   Appendix \ref{app:X}.
 
 The third diagram involves the $h_{(3)}  h_{(3)} \ovp \vp $ vertex required for the manifest covariance of the scalar action 
 under the spin 3 gauge transformations. 
 As discussed above, its  contribution is expected to vanish on a Bach-flat ({\em e.g.}  Einstein) 
 space. We shall impose, for simplicity,  the  condition that $g_{ab} + h_{ab}$   is Einstein, 
 which (for the TT field $h_{ab}$)  amounts to the condition $\Box h_{ab}=0$   to be assumed below.

  The explicit results for the contributions of  the  diagrams  (a) and (b) are  given  by 
(\ref{F.2}) and (\ref{F.3}). The total action $\text{S}_{233}$  turns out to be consistent  with 
the   result   of \cite{Nutma:2014pua}  for the linear in curvature term  in the  $\text{S}_{33}$ action
where it  was found by demanding  the spin 3 gauge invariance 
 to the leading order   in   curvature expansion.\foot{We have used  the Mathematica 
 notebook provided in  \cite{Nutma:2014pua} to  check  that they result 
$\text{S}_{233}^{\rm NT}$ is equivalent to ours $\text{S}_{233}$  for a particular choice of 
free parameters  in  \cite{Nutma:2014pua}  associated with field redefinitions, 
 total derivatives, and  use of 4d  identities   and after       restricting to TT fields   and imposing
 the  linearized  Einstein space constraint $\Box h_{ab}=0$.}
 As a simple  illustration of the agreement,   let us   formally set $h_{ab}$ to be constant ({\em i.e.} ignore all curvature terms). Then  
\begin{align}
\la{4.22}
\mathscr L_{233}^{\rm NT} &= - \tfrac{135}{7}\big( h^{ab} h^{cde} \
\partial_{ab}\Box^{2}h_{cde} 
 +h^{ab} h_{a}{}^{cd} \
\Box^{3}h_{bcd}\big) = \mathscr L_{233}= \mathscr L_{233}^{(a)}+ \mathscr L_{233}^{(b)}  \ ,   \\
\mathscr L_{233}^{(a)} &= - \tfrac{135}{7}\,\big( h^{ab} h^{cde} \
\partial_{ab}\Box^{2}h_{cde} - h^{ab} h_{a}{}^{cd} \
\Box^{3}h_{bcd}\big),\qquad \ \ \ \ 
 \mathscr L_{233}^{(b)} =-\tfrac{270}{7}\, h^{ab} h_{a}{}^{cd} \
\Box^{3}\,h_{bcd} \ ,  \no  
\end{align}
where we  made a  field rescaling to account for  a  difference  in our choice of normalizations 
compared
to  \cite{Nutma:2014pua}.\foot{The agreement of the actions for constant $h_{ab}$   is,  in  general,  guaranteed 
if  the flat space actions  match  (as one can generate  a constant metric by a coordinate redefinition) 
but it still provides a formal consistency check of the two results.}

\section{Spin 1--3 mixing term contribution to   UV divergences} 
\la{sec:CC}

Starting with the induced action for the  tower of CHS fields  
one may  attempt to compute  the corresponding   UV divergences and thus conformal anomalies. 
Assuming we expand near the vacuum  point (with Bach-flat  metric) 
so that all  linear terms in \rf{310} 
vanish, the action  will begin  with  the quadratic term $\SS^{(2)}$ in \rf{3100}
and thus  the 1-loop correction to the CHS partition function  will be  
  expressed  in terms  of   determinants of  the operators $\OO_{s,s'}$  in \rf{3100}. 
  
  In general, the   logarithmic divergences  or  conformal anomalies in  curved $\dd=4$  background 
     are  governed by the   coefficients a  and c in the corresponding  Seeley coefficient  (see, {\em e.g.}, \cite{Duff:1993wm})
\begin{align}
\la{55} 
\G = -\log Z_{\text{CHS}} = & 
- \tfrac{1}{(4\pi)^{2}} \log \Lambda_{_{\rm UV}}\ \int d^{4}x\,\sqrt{g}  \ b_4 (x) \ \  + \ \ \text{finite} \ , \\
 \la{555}
b_4= &
 -\text{a}\,R^{*}\,R^{*}+\text{c}\,C^{2}  \ . 
\end{align}
To extract  the coefficients a  and c   
one  may  compute 
$b_{4}$   separately    in a  conformally flat Einstein  background  where
 $b_{4}=-\text{a}\,R^{*}R^{*}$   and in a  Ricci flat   background  
 where ${b_{4} = (\text{c}-\text{a})\,C^{2}}$.
 
In the conformally flat case the CHS kinetic operators  are  diagonal in spin and 
  factorize into products of  second-derivative operators \ci{Tseytlin:2013jya,Metsaev:2014iwa,Nutma:2014pua}
  and thus the corresponding a-anomaly coefficient can be computed \cite{Tseytlin:2013jya}
   using    standard methods like in \ci{Christensen:1978md}\foot{This was  
 done  also  using  the  AdS$_5$ related  method 
   \ci{Giombi:2013yva}.}
   \be
\la{5.33}
\text{a}_{s} = \tfrac{1}{720}\,\nu_{s}\,(3\,\nu_{s}+14\,\nu_{s}^{2})\ , \qquad\qquad 
 \nu_{s}\equiv s(s+1)  \ . 
\ee
In the special cases of  conformal  spins  $2$ ({\em i.e.} Weyl graviton) 
and $3/2$ (conformal gravitino)  this factorization  on (A)dS$_4$   or $S^4$   background was observed  long ago in 
\cite{Tseytlin:1984wj,Fradkin:1983zz,Deser:1983mm,Deser:1983tm}.

The factorization  of the  Weyl graviton and conformal gravitino   kinetic operators 
 turns out to   hold  also  in a Ricci-flat background  \cite{Fradkin:1983zz,Fradkin:1985am}. 
In  \cite{Tseytlin:2013jya} it was  conjectured   that this   factorization  may  apply 
to all   CHS   kinetic operators in $R_{ab}=0$    background 
 leading to the following prediction for the spin $s$   contribution to  the  c-coefficient in \rf{555} 
\be
\la{5.3}
\text{c}_{s} = \tfrac{1}{720}\,\nu_{s}\,(4-42\,\nu_{s}+29\,\nu_{s}^{2})  \ . 
\ee
As was argued  in \cite{Nutma:2014pua},  the   Ricci-flat factorization  conjecture 
may  not be true  in general  for $s > 2$  as there   should be  curvature derivative dependent  terms   like $\nabla^n R_{....}$ 
that represent   obstructions  to factorization. 
However,  such terms  can not  contribute to the 
UV divergences  \rf{55},\rf{555}   and thus  to the value of $\rc_s$  on dimensional grounds.
Note also that  the general argument  in 
\cite{Nutma:2014pua}  was   under the assumption    that the CHS kinetic operator $\OO_{2s}$  is  diagonal and
  gauge invariant  separately  for each $s$, which is not,  in general,  true as we have seen  above.
  Still,  even ignoring   such derivative terms  the factorization conjecture for $s\geq 3$ remains to be proved.

Regardless the   validity of the  factorization conjecture, what was not included   in the previous   analysis  is a
potential contribution to \rf{55},\rf{555} coming from   non-diagonal mixing terms like   $\SS_{13}$ in \rf{4.1},\rf{4.12}. 
Such mixing terms  vanishing in conformally flat Einstein  background  can not contribute to anomaly a-coefficient but may contribute to c-coefficient. 
Here we   will    concentrate  on  the spin 1--3   sector discussed above.  
For   $s=1$  (Maxwell)  field  we have the standard result  $\rc_1 = {1\ov 10}$  ($\nu_1=2$ in \rf{5.33},\rf{5.3}) 
while  for the diagonal  $s=3$  contribution  (assuming factorization of $\OO_6$)   we expect  from \rf{5.3} to get 
 $\rc_3 = {919\ov 15}$  ($\nu_3=12$).
 
\begin{figure}[t]  
  \centering
  \ifLocalFigs
 \includegraphics[scale=0.4]{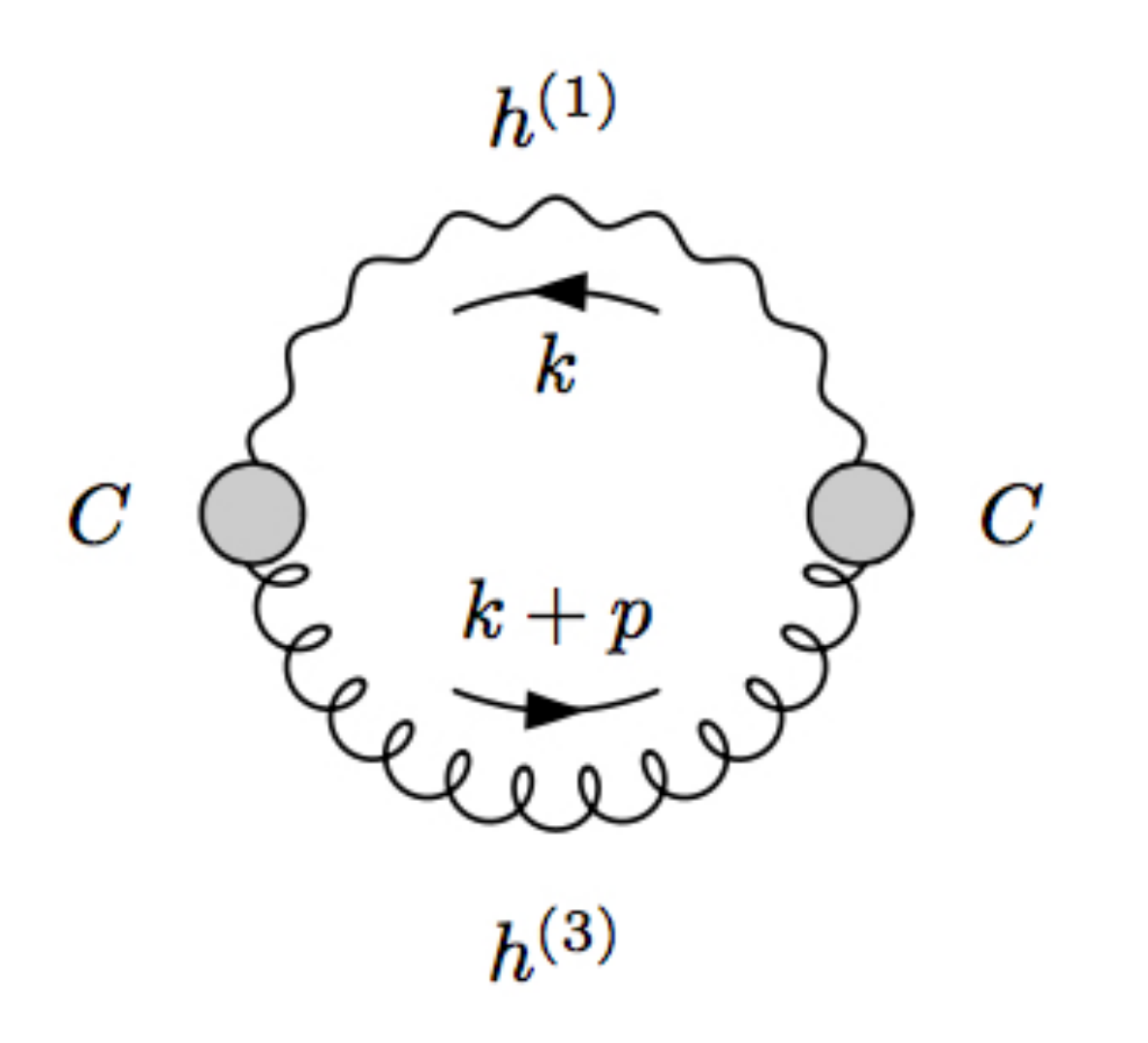}
 \else
 \includegraphics[scale=0.4]{./Figures/FeynArts/CC.pdf}
 \fi
   \caption[]
   {One-loop diagram in the spin 1-- spin 3    theory 
   contributing to the $C^{2}$  UV  divergence. 
  \label{fig:cc}  }
\end{figure}

  Let  us  consider the background metric to be generic  (not necessarily Einstein). 
  The mixed 1-3 term in the induced action (\ref{4.12})   contains   the  $ C \nabla  h_{(1)}  \nabla h_{(3)}$   and 
  $  (\nabla  R) (  \nabla  h_{(1)} ) h_{(3)}  $ 
  vertices    while the kinetic terms   are $  h_{(1)}  (\na^2 + ...)  h_{(1)}    +   h_{(3)}  (\na^6 + ...)  h_{(3)} $. 
  It is   easy to see on dimensional grounds that   only   the first $ C \nabla  h_{(1)}  \nabla h_{(3)}$   mixing vertex 
   may  in principle contribute to the $C^2$  UV divergences in \rf{555}. We may thus start   directly with the  
   simple  quadratic action \rf{4.17}.
   The corresponding  additional   $C^2$  contribution   may  come   from 
    the   UV divergent   part of the diagram in  Fig.~\ref{fig:cc}.
As the mixing vertex contains already one   factor  of the Weyl tensor,  to  find this contribution 
it is sufficient to consider the flat-space spin 1 and spin 3 propagators in TT gauges. 
 We  
   get  for the resulting contribution to  the  effective action in momentum representation 
\begin{align}
\no
\Gamma (p) = &{\te \frac{1}{2}}\,  8^{2}\, n_{1}\, n_{3}\,\int\frac{d^{\dd}k}{(2\,\pi)^{\dd}}  
 \,C\indices{^{aced}}(-k_{e}-p_{e})(k_{a}g_{bq}-k_{b}g_{aq}) \\
& \qquad  \times C\indices{^{a'c' e'd'}}
(k_{e'}+p_{e'})(-k_{a'}g_{b' q'}+k_{b'}g_{a' q'})
\frac{P^{q , q'}(k)}{k^{2}}\frac{P\indices{^{b}_{cd, }^{b'{ c' d'}}}(k+p)}{[(k+p)^{2}]^{3}}\ ,  \la{5.4}
\end{align}
 where $n_{s}$  are normalizations   in  the   flat-space CHS quadratic actions 
\be \la{5.5} 
\frac{1}{2\,n_{s}} \,\int d^{4}x\, h_{a_{1}\dots a_{s}}\, P_{ b_{1}\dots b_{s}  }  ^{ a_{1}\dots a_{s}  }  \Box^{s}\,h^{b_{1}\dots b_{s}} \ ,  \ee
while $P_{a, b}(k)$ and $P_{abc, def}(k)$ are the  spin 1 and spin 3  TT projectors  $P_{(s) (s')}$ 
 in momentum representation 
(see, {\em e.g.}, eq.(3.6) and footnote 18 in
\cite{Beccaria:2016syk}).  Then after  some standard manipulations 
(introducing Feynman parameters  and  shifting loop momentum)  we get 
\footnote{
We thank S. Nakach for correcting the overall coefficient in (\ref{5.6}) in the first version of this paper.
}
\be
\la{5.6}
\Gamma (p) = 
\frac{224}{5}
\, n_{1}\, n_{3}\,C_{a}{}^{e}{}_{b}{}^{f} (p)   \ C_{cedf} (-p) \int^1_0 d x  \int\frac{d^{\dd}k}{(2\pi)^{\dd}} \frac{k^{a}\,k^{b}\,k^{c}\,k^{d}}{\big[k^{2}+M^{2}(p,x)\big]^{4}}+\text{finite}.
\ee
The  UV  divergent part  is then ($\eps=4-\dd\to 0$)\foot{We use that one can make  the replacement 
  $k_{a}k_{b}k_{c}k_{d} \to \frac{1}{24}(g_{ab}\,g_{cd}+g_{ac}\,g_{bd}+g_{ad}\,g_{bc})\,(k^{2})^{2},
$
and the standard integral:
$$
\int  \frac{d^\dd k}{(2\pi)^\dd}\frac{(k^{2})^{a}}{(k^{2}+M^{2})^{b}} = 
\frac{\Gamma(b-a-\dd/2)\Gamma(a+\dd/2)}{(4\pi)^{\dd/2}\Gamma(b)\Gamma(\dd/2)}\,(M^2)^{\dd/2+a-b}.
$$
Note that in dimensional regularization  in \rf{55} one has  $\log \Lambda_{_{\rm UV}} \to - {1\ov \eps}$. }
\be
\la{5.9}
\Gamma_{_{\rm UV}}  = \frac{28}{5}\,\frac{
\ n_{1}\,n_{3}}{(4\pi)^{2}\,\eps}\,\,C_{abcd}C^{abcd}\  . 
\ee
Our normalizations of the flat-space currents in \rf{22},\rf{2.2}  correspond to 
$n_{1}=3$   and  $n_{3}=\frac{7}{90}$  in \rf{5.5} (see  \rf{4.17}, \rf{4.14})
so that the final  result for the contribution of the 1-3  mixing to the coefficient c in \rf{555} is 
\be
\la{5.11}
\rc_{13}  =  \te 
\frac{98}{75}  \ .
\ee
\iffa 
\footnote{
\la{fn}
In the normalization of  \cite{Beccaria:2016syk}, the source term for a TT spin-$s$ field is
$\widehat{h}_{a(s)}\,J^{a(s)}=\frac{i^{s}}{s!}\,\widehat{h}_{a(s)}\,\ovp\,\partial^{a(s)}\,\vp$ and it leads to the quadratic
Lagrangian $\frac{1}{2^{s+1}\,(2s+1)!}\,\widehat{h}_{a(s)}\,\Box^{s}\,\widehat{h}^{a(s)}$. For spin 1, we have from (\ref{F.1})
$h_{a}J^{a} = 2\,i\,h_{a}\,\ovp\partial^{a}\vp$ so $2h=\widehat{h}$ and indeed 
$\frac{1}{2^{1+1}\,(2+1)!}\,\widehat{h}_{a}\,\Box\,\widehat{h}^{a} = \frac{1}{6} h_{a}\Box h^{a}$
in agreement with $-\frac{1}{12}F^{2}$ in (\ref{G.10}). For spin 3, from (\ref{K.12}), we have 
$h_{abc}\widetilde J^{abc} = 120\,i\,h_{abc}\,\ovp\partial^{abc}\vp$, so now $120h = 
-\frac{1}{6}\,\widehat h$
and the quadratic term is $\frac{1}{2^{3+1}\,(2\cdot 3+1)!}\,\widehat{h}_{a}\,\Box^{3}\,
\widehat{h}^{a} = \frac{45}{7}\,h_{abc}\Box^{3}h^{abc}$.
} and 
arrive at the final divergent contribution 
\be
\la{5.10}
\frac{1}{16\,\pi^{2}\,\eps}\,\int d^{4}x\,\sqrt{g}\ \ \frac{392}{5}\,C_{abcd}C^{abcd}.
\ee
\fi
The total contribution to c from the   1-3 sector is thus
$\rc_{1+3} = \rc_1 + \rc_{13} + \rc_{3} = 
{1\ov 10}  + \frac{98}{75} + {919\ov 15}
$, {\em i.e.}  
the magnitude of the mixed term  contribution is intermediate between 
the pure spin 1 and 3 ones.

 In general, other  similar  higher spin  mixing terms  are   expected to appear and thus should 
 also contribute to the $C^2$   UV divergence.   
 Indeed, just on dimensional grounds   one  may have   for spin $s$ and   spin $s'$  
 part of the   CHS action expanded  near flat space 
 (we assume  the  fields to be TT and ignore normalization factors, {\em cf.} \rf{4.17})\foot{To determine the
  mixing term  here  requires the computation of the $h_{(2)} 
 h_{(s)} h_{(s')}$  term in the CHS   action in a  near flat space expansion.} 
 \be \la{5.111}
\!\!\! \!  \SS_{s+s'}  = \int d^4x  \Big[  h_{(s)} (\del^{2s}  + ...)  h_{(s)} + h_{(s')}( \del^{2s'} + ...)  h_{(s')}   +  
 \sum_{n} C\ \del^n  h_{(s)}  \del^{n'} h_{(s')} + ...\Big]    
  \ , 
 \ee 
 where  $C$  stands for the Weyl tensor  and $n+n'=s+s'-2$  to balance dimensions. 
 Then the  UV  singular   $C^2$ contribution to the one-loop effective  action is proportional to $\int d^4 k { k^{2n+2n'} \ov k^{2s+2s'} } \sim \int  {d^4 k \ov k^4}$. {\em i.e.} 
 is logarithmically divergent  and thus contributes to c-coefficient in \rf{555}. 

  It  then   remains  an open question if these   mixing term     contributions  may  change the expectation 
 \ci{Tseytlin:2013jya}  that the  regularized 
  sum of  all CHS contributions to  the  conformal anomaly   c-coefficient should   vanish, like  that happens for the total a-coefficient  
  coefficient 
  \ci{Giombi:2013yva,Tseytlin:2013jya}.


\section*{Acknowledgments}
We are grateful to  R. Roiban for a collaboration at an  initial stage and useful comments on the draft. 
We thank  A.  Barvinsky, M. Grigoriev, S. Kuzenko, R. Metsaev,  H. Osborn  and M. Taronna  for  useful 
discussions. We also thank M. Taronna  for  his kind clarifications  of the computation  reported  in 
 the Mathematica notebook in \cite{Nutma:2014pua}.
The work of AAT  was   supported by the ERC Advanced grant no. 290456,
 the  STFC Consolidated grant ST/L00044X/1, the ARC  project DP140103925 
  and   the Russian Science Foundation grant 14-42-00047 at Lebedev Institute.

 \section*{Appendices} 
 
\appendix

\section{Curvature   identities}
\la{app:bach}

In four dimensions,  one has  the  following useful identities  for the Weyl tensor 
\be
\la{A.1}
C^{acde} C^{b}{}_{cde} =  \tfrac{1}{4} g^{ab} \
C_{cdef} C^{cdef}, \qquad\qquad 
C^{acde} C^{b}{}_{dce} =  \tfrac{1}{4} g^{ab} \
C_{cedf} C^{cdef}.
\ee
The Bach tensor is defined by 
\begin{align}   \ \la{A.2}
\!\!\!B_{ab} = &-\tfrac{1}{4}\,\tfrac{1}{\sqrt{g}}\,\tfrac{\delta}{\delta g^{ab}}
\,\int d^{4}x\,\sqrt{g} \, C^{2}_{abcd}  \\
=& \ R_{a}{}^{c} R_{bc} -  \tfrac{1}{3} R_{ab} R +   R^{cd} \ C_{acdb}   -
\tfrac{1}{2} \nabla^2 R_{ab} +  \tfrac{1}{6} \nabla_{a}\nabla_{b}R
    -  \tfrac{1}{4} g_{ab} (R_{cd} R^{cd} - \tfrac{1}{3} R^2 - \tfrac{1}{3} \nabla^2 R ) \no
\end{align}
Introducing the  
Schouten  tensor $P_{ab} = \frac{1}{2}(R_{ab}-\frac{1}{6}\,
R\,g_{ab})$,  the Bach tensor  may be written as 
\begin{align}
\la{A.3}
 B_{ab} &= \nabla^{c}\nabla_{a}\,P_{bc}
-\nabla^{2}\,P_{ab}+P^{cd}\,C\indices{_{a cd b} } = 
\big( \nabla^{c}\nabla^{d}+
  \tfrac{1}{2} R^{cd}\big) \,C_{acdb}\ ,
\end{align}
where the second equality follows from the  Bianchi identities.\foot{Let us note that some   papers have different signs 
appearing in the expression for the Bach tensor which is  usually  related to 
 different conventions for the sign of the  curvature.}
 
\section{Vanishing of spin 3 linear term  in  induced action}
\la{app:linear}

Let us consider the  linear term in the induced action \rf{310},\rf{3100}
\be 
\la{b1} 
\SS^{(1)}_3= \int d^{4}x\,\sqrt{g}\, h^{abc}\,B_{abc}  \ , \qquad \qquad B_{abc}= \langle  J_{abc} \rangle_{_{\rm UV}}\ , 
\ee 
 given  by the coefficient of the logarithmic divergence in the 1-point function of the spin 3 current \rf{2.18}.
 Since the UV singular part of the  1-point function of spin 1  current is equal to zero,  $B_a =  \langle  J_{a} \rangle_{_{\rm UV}}=0$, 
 the   relation \rf{2.20}  implies that  dimension 5 tensor   $B_{abc}$ should be covariantly conserved, 
 $\nabla^a B_{abc} =0$.
  In addition, it should   parity-even (as $J_{abc}$ in \rf{2.18}  and $S_0$ in \rf{2.7} are)   
   and Weyl-covariant with weight  -2  so that   $\SS^{(1)}_3$ is Weyl-invariant  ({\em cf.} \rf{2.9}). 
  
Examining the most general  candidates for $B_{abc}$  satisfying these conditions  we did not find any solutions, 
i.e. we should have 
\be    B_{abc} =0 \ . \la{b2}  \ee
In \cite{Grigoriev:2016bzl}  it was suggested that $B_{abc} $ may be proportional to Eastwood-Dighton tensor
\be \la{b3} 
E_{abc}= C^{ef}_{\ \   cb} \na^d C^*_{defa} -  C^{* ef}_{\ \ \ \  cb} \na^d C_{ defa}  \ ,   \qquad \qquad C^*_{abcd} = \ha  \eps_{abef}  C^{ef}_{\ \   cd} 
\ , \ee 
which  satisfies the Weyl-invariance and conservation conditions  and vanishes on conformally Einstein spaces.
 However,  this tensor is parity-odd
and thus cannot appear in the expectation value $\langle  J_{abc} \rangle_{_{\rm UV}}$.

\iffa 
D_a D_b D_c   C_kmnp we need to contract two indices    and symmetrise
the rest  to get K_pqr ; that means, modulo commutators   giving CC
terms that we should   have
D_a  D_m D_n C_b mn c
But D_m D_n C_b mn c   = 0 modulo R C  term if we use  Bach-flatness.
Here Weyl-invariance was not used
(and D_a  D_m D_n R_b mn c  is same  as D_a  D_m D_n C_b mn c
as Ricci terms vanish or   traces).
Instead, you  rule out D_a  D_m D_n C_b mn c
by Weyl invariance condition.
We may say this equivalently:
D_(a  B_bc)  is not Weyl invariant so  cannot appear.
This is  easy to show directly -- that this is not Weyl inv -- as B_bc
has definite Weyl weight.
\fi

Let us note that  in the special  case  of   an  Einstein space   background  $R_{ab} = {1\ov 4}  R g_{ab}$ 
we do not need the condition of Weyl invariance to show that  $B_{abc}=0$. Indeed, the dimension 5 tensor 
$B_{abc}$  must   be constructed from the 
 Weyl tensor and covariant derivatives,   {\em i.e.}  
 it should be   of the form $\nabla CC$ or $\nabla^{3}C$ or explicitly 
 (ignoring  $g_{ab}$-terms that decouple upon contraction with traceless $h_{abc}$)  
\begin{align}
\la{B.1}
B_{abc} = k_{1}\,\nabla^{e}\nabla^{d}
\nabla_{c}C_{adbe}  +  k_{2}\,C_{a}{}^{def} 
\nabla_{c}C_{bdef} + k_{3}\,C_{a}{}^{def} \
 \nabla_{c}C_{bedf}
 +   +  k_{4}\,
C_{a}{}^{def}  \nabla_{f}C_{bdce} \ . 
\end{align}
The first term  can be   related  to the  last  three as   for 
an Einstein space 
\be
\la{B.3}
\nabla^{e}\nabla^{d}\nabla_{c}\,R_{adbe} = 
- R_{b}{}^{def} \nabla_{d}R_{aecf} + R_{c}{}^{def} \nabla_{f}R_{adbe} \
-  R_{a}{}^{def} \nabla_{f}R_{becd} \ . 
\ee
The  second and third  terms  do not contribute to $ h^{abc}\,B_{abc}$    because of the identities like  ({\em cf.} \rf{A.1}) 
\be
\la{B.2}\te 
C_{a}{}^{def} \nabla_{c}C_{bdef} = \frac{1}{2}\nabla_{c}
(C_{a}{}^{def}C_{bdef}) =  \frac{1}{8}\,g_{ab}\,\nabla_{c} C^{2} \ . 
\ee
Using the  Bianchi identity  we  also  have 
$
C_{a}{}^{def} \nabla_{f}C_{bdce} = - C_{a}{}^{def} \nabla_{b}C_{cedf} -  
C_{a}{}^{def} \nabla_{d}C_{becf} .
$
These terms  do not contribute upon contraction with a totally   symmetric traceless $h_{abc}$. 

One may also study  more general linear in spin 3  terms which also involve the  vector 
 field strength $F_{ab}$. One finds that the only possible 
  term   with one power of  $F_{ab}$
   is proportional to the combination  appearing in (\ref{4.12})   that we obtained 
 by the direct computation of the induced action. 
The   only term quadratic in  $F_{ab}$   and  linear in $h_{abc}$  must   
be (on dimensional and covariance grounds)  proportional to 
$F_{ab}F\indices{^{b}_{c}} \nabla_{d}\,h^{acd}$.
Such term   is not,   however,  Weyl invariant
and  thus   can   not appear in the induced action.

\section{Background covariant  computation of  UV pole  parts\\ of correlators of currents}
\la{ab}

In this Appendix  we shall briefly  explain  the strategy  of the computation
of the   UV pole (logarithmic divergence) part  of the 
current-current  correlators   like   
$\langle J_{(s)} \,J_{(s')}\rangle_{_{\rm UV}}$   appearing in  (\ref{4.2}). 
We shall follow the approach and notation of  \cite{Jack:1983sk}.

Starting with  the explicit expression for  the  bilinear currents  like in \rf{2.10},\rf{2.12},\rf{2.18} 
one may  express   the  correlator at separated points $\langle J_{(s)}(x)\,J_{(s')}(x')\rangle$
in terms of the curved space scalar propagators  $G(x, x')$  getting   sum of terms like 
\be
\la{X.1}
\nabla^{n}\,\nabla'^{n'}\,G\ \ \nabla^{m}\, \nabla'^{m'}\,G  + ... ,\qquad \qquad n+m=s, \ n'+m'=s',
\ee
where dots stand  for  other potential 
 terms with less covariant derivatives  but extra factors of curvature and its derivatives. 
For our purpose of extracting  the UV singular part of the correlator it is sufficient to  keep only the  part of $G$  
which is most 
singular in the  coincidence $x\to x'$   limit 
\be
\la{X.2}
G(x,x')  \to  \frac{\Delta^{1/2}(x,x')}{8\,\pi^{2}\,\sigma(x,x') } \ ,
\ee
where $\sigma(x,x')$ is  half the geodesic distance between $x$ and $x'$ and  $\Delta$ (not to be confused with the 
 Laplace   operator) 
 is defined by 
(see, e.g.,  \cite{Christensen:1976vb} for a detailed  discussion of  properties of these bitensors)
\be
\la{X.3}
{ \Delta} \equiv  |g|^{-1/2}\,|g'|^{-1/2}\,\det D_{ab'}\ ,\qquad  \qquad D_{ab'}(x,x') = -\frac{\partial^{2}}{\partial x^{a}\,
\partial x^{' b}}\,\sigma(x,x').
\ee
Using   (\ref{X.2}) in (\ref{X.1}), we obtain  terms whose denominator is a power of $\sigma$
while the numerator is a tensor that involves covariant derivatives of $\sigma$ and $\Delta$.

To find the UV singular  part  of   such terms,  
in the dimensional regularization approach of \cite{Jack:1983sk}, one  is to make the following 
replacement  ($\eps= 4-\dd$) 
\be
\la{X.4}
\frac{1}{\sigma^{2}}\to \frac{8\,\pi^{2}}{\eps}\,\delta^{(4)} (x,x')\  ,
\ee 
where $\delta^{(4)}(x,x')$ is the biscalar curved space $\delta$-function, 
 $\int d^{4}x\,\sqrt{g} \ \delta^{(4)}(x,x') = 1 $. 
 Terms with  powers of $1/\sigma$ higher than 2 
should  be  first reduced to $1/\sigma^2$ terms  by iterative  use  of  the   identity 
\be
\la{X.5}
\frac{1}{\sigma^{p+1}} = \tfrac{1}{2\,p\,(p-1)}\,\Delta^{-1/2}\,
(\nabla^{2}-Y)\,\frac{\Delta^{1/2}}{\sigma^{p}}\ ,\qquad \qquad Y=\Delta^{-1/2}\,\nabla^{2}\,\Delta^{1/2}.
\ee
Use of \rf{X.4}   then  gives  terms   
 with powers of the differential operator in the r.h.s. of (\ref{X.5}) acting on $\delta^{(4)}(x,x')$.
We  may then  use  integration  by parts  and  take the 
coincidence limit $x\to x'$.

 This last step is non trivial because the coincidence limit 
does not commute with covariant derivatives of the biscalars $\sigma$ and $\Delta$.\footnote{Besides, one has to deal 
with the technical problem of separating covariant derivatives at $x$ and $x'$. This may be done
systematically by exploiting Synge's theorem 
and its multi-index generalization proved in \cite{Christensen:1976vb}.} 
This may be automatized to provide a table of substitution rules. Denoting by a square
bracket the coincidence limit, the simplest  examples  are   
\begin{align}
& [\sigma]=0\ , \qquad  \ [\nabla_{a}\sigma]=0\ , \ \qquad  [\nabla_{a}\nabla_{b}\sigma]=g_{ab}\ , \ \qquad  
 [\nabla_{a}\nabla_{b}\nabla_{c}\sigma]=0\ , \no\\
&[\nabla_{a}\nabla_{b}\nabla_{c}\nabla_{d}\, \sigma] = -\tfrac{1}{3}(
R_{dbca}+R_{dacb})\ , \ \qquad  \dots   \la{X.6} \\
& [\Delta^{1/2}]=1\ , \qquad  \ [\nabla_{a}\Delta^{1/2}]=0\ ,  \ \qquad   [\nabla_{a}\nabla_{b}\Delta^{1/2}] = \tfrac{1}{6}\,R_{ab}\ , \no \\
 &[\nabla_{a}\nabla_{b}\nabla_{c}\, \Delta^{1/2}]= \tfrac{1}{12}\,(
\nabla_{a}R_{bc}+\nabla_{b}R_{ac}+\nabla_{c}R_{bc}), \qquad \dots   \la{X.7}
\end{align}
The substitutions (\ref{X.6}) and (\ref{X.7})  
produce a non trivial dependence on the background curvature. 
When  the correlator is contracted with $h_{(s)}$ fields, i.e. $h^{(s)} \langle J_{(s)} \,J_{(s')}\rangle_{_{\rm UV}}\,  h^{(s')} $, 
the integration by parts   mentioned above will produce  terms with derivatives acting on   higher  spin fields.

\section{Stress tensor of the free   spin 3  field  in flat space  
}
\la{app:T3}

Here   we shall   comment on the  special   structure of the   spin 3   flat space kinetic term 
 in (\ref{4.14})  and its stress tensor. 
   Let us  start with  the general 3-parameters 
  Lagrangian 
\be
\la{C.1}
{\mathscr L} = h^{abc} \Box^{3}\,h_{abc} +  k_{1}\, h^{abc}\, \partial_{abcdef}\,h^{def} 
+ k_{2}\,  h^{abc} \Box\,
\partial_{bcde}\,h_{a}{}^{de}  +k_{3}\,h^{abc}\, \Box^{2}\,
\partial_{cd}\,h_{ab}{}^{d} \  ,
\ee
and look for a  {symmetric  traceless}
stress tensor $T_{ab}\sim h_{(3)}\partial^{6}h_{(3)}+\cdots$ 
which is conserved and gauge invariant on  the equations of motion following   from  (\ref{C.1}).
There are 254 possible structures  in  such  $T_{ab}$, {\em i.e.} $
(\partial^{n}h\,\partial^{m}h)_{ab}\quad \text{or}\quad g_{ab}\,
(\partial^{n}h\,\partial^{m}h)$ 
with $ n+m=6$.
The $T_{ab}$    with required properties is   found  only if $\mathscr L$ in (\ref{C.1}) is proportional to
$\mathscr L_{33}^{\rm flat}$ in (\ref{4.14}).\footnote{We 
omit trivial improvement terms not contributing to the conserved 
charges and obeying all the requirements automatically, {\em i.e.} without using the equations
of motion.} 
To show  the converse  
requires an explicit calculation   which gives 
\begin{align}
\la{C.3}
T^{ab}= &\tfrac{1}{2} h^{bcd} \partial^{a}{}_{cdefp}h^{efp} - 2 \
h^{bcd} \partial^{a}{}_{def}\Box h_{c}{}^{ef} + 
\tfrac{5}{4} h^{bcd} \
\partial^{a}{}_{e}{} \Box^{2} h_{cd}{}^{e} \nonumber \\ 
& + \tfrac{1}{2} h^{acd} \partial^{b}{}_{cdefp}h^{efp} - 2 \
h^{acd} \partial^{b}{}_{def}\Box h_{c}{}^{ef} + 
\tfrac{5}{4} h^{acd} \
\partial^{b}{}_{e}\Box^{2} h_{cd}{}^{e} \nonumber \\ 
& -  h^{bcd} \partial_{cdef}\Box h^{aef} -  h^{acd} \
\partial_{cdef}\Box h^{bef} + h^{abc} \
\partial_{cdef}\Box h^{def} + \tfrac{5}{2} h^{bcd} \
\partial_{de}\Box^{2} h^{a}{}_{c}{}^{e} \nonumber \\ 
& + \tfrac{5}{2} h^{acd} \
\partial_{de}\Box^{2} h^{b}{}_{c}{}^{e} -  \
h^{abc} \partial_{de}\Box^{2} h_{c}{}^{de} -  
\tfrac{5}{4} h^{bcd} \
\Box^{3} h^{a}{}_{cd} \nonumber \\ 
& -  \tfrac{5}{4} h^{acd} \
\Box^{3} h^{b}{}_{cd}
 +\eta^{ab}\,\Big(- \tfrac{1}{4} h^{cde} \partial_{cdefpa}h^{fpa} + \tfrac{3}{2} \
h^{cde} \partial_{defp}\Box h_{c}{}^{fp} \nonumber \\
& -  \tfrac{15}{8} h^{cde} \
\partial_{ef}\Box^{2} h_{cd}{}^{f}  + \tfrac{5}{8} h^{cde} \
\Box^{3} h_{cde}\Big).
\end{align}
Here  the $\eta^{ab}$ term is  proportional (up to a total derivative) to $\mathscr L_{33}^{\rm flat}$ in \rf{4.14}, 
as  expected on general grounds.

The fact that \rf{C.3}   is not manifestly invariant under spin 3 gauge transformations 
may  be compared with the   lower spin cases. 
In the  spin 1 case  
$T_{ab} = F_{ac}F\indices{^{c}_{b}}-\frac{1}{4}g_{ab}F^{2}$ is traceless,  symmetric and
conserved on the equations of motion  but also manifestly gauge-invariant. 
The latter feature   does not  automatically generalize to higher  conformal spins.
 Let  us  consider  the  $s=2$ case   and expand the  Weyl gravity  action near a generic metric $g_{ab} \to g_{ab} + h_{ab}$ 
 (with traceless $h_{ab}$) 
\begin{align}
\la{C.4}
S(g; h) &={\te{ 1\ov 4}}  \int d^{4}x\, \sqrt{g}\, C^{2}_{abcd}(g+h) = 
\int d^{4}x\,\sqrt{g} \big [ \ {\te{ 1\ov 4}}  C^{2}_{abcd}(g)+ B(g)\, h +\tfrac{1}{2} h\, \mc{O}_{4}(g)\, h +\dots\big] \notag \\
&= S(g) + S^{(1)}(g; h)+S^{(2)}(g; h)+\dots \ , 
\end{align}
where  $B$ is  the Bach tensor. 
Under  the gauge variation  
$\delta_\eps h_{ab} = \nabla_{(a} \eps_{b)}-\frac{1}{4}g_{ab}\,\nabla^{c}\eps_{c}$,
\begin{align}
\la{C.5}
0 = S(g; h+\delta_\eps h) -S(g; h) = \int d^{4}x\,\sqrt{g} B(g) \, 
\delta_\eps h +S_{2}(g; h+ \delta_\eps  h)-S_{2}(g; h)+\dots.
\end{align}
Applying $\delta/\delta g_{mn}$ and replacing  the background metric by the flat one,  $g_{ab}\to \eta_{ab}$, gives
\begin{align}
\la{C.6}
 T_{cd}(h+\delta_\eps h)-T_{cd}(h) = -  \Big[ \frac{\delta B^{ab}}{\delta g_{cd}}  \delta_\eps h_{ab} \Big]_{g_{ef}= \eta_{ef}} 
  -  \Big[ \frac{\delta S^{(2)}}{\delta h_{ab}}\, \frac{\delta(\delta_\eps h_{ab})}{\delta g_{cd}}\Big]_{g_{ef}= \eta_{ef}}+\dots \ , 
\end{align}
where  $T_{cd} $ is the stress tensor defined  as the  variation of the action over the metric.
The  first term on the r.h.s. then vanishes as $B_{ab}$  is covariant; the second term  vanishes on 
the the equations of motion of $h_{ab}$. 
Thus  $T_{cd}$ is gauge invariant only on the equations of motion, in contrast to the spin 1 case.


\section{Spin 3 gauge invariance  in Einstein  background}
\la{app:check-gauge}

Here  we provide some details of the check of the invariance of the  spin 1 plus  mixed spin 1-3 term in  the quadratic action \rf{4.17}
in an Einstein background  under the  spin 3   gauge  transformation  in \rf{4.16}. 
Explicitly, we consider 
\begin{align}
\la{D.1}
& \delta\int d^{4}x\,\sqrt{g}\big(- {\te \frac{1}{12}}F^{2}_{ab}
+8\, C^{abdc}\,F_{ap}\,\nabla_{c}h_{bd}^{ \ \ p}\big) \equiv  \int d^{4}x\,\sqrt{g}\,(Q+Q'), \notag \\
&Q = h^{a}\,\Delta_{a,pq}\,\eps^{pq}, \qquad Q'=h^{abc}\,\Delta'_{abc,pq}\,\eps^{pq},
\end{align}
where $\delta$ acts   according to  \rf{4.16} and 
  $\Delta_{a,pq}$, $\Delta'_{abc,pq}$ are
differential operators  containing  $\nabla_a$  and   $\overleftarrow{\nabla}_a$. 
We want to show that 
the  part of the   variation   not depending on $h_{abc}$, {\em i.e.} $Q$,  vanishes.
The cancellation  of $Q'$ requires adding the variation of the  last quadratic spin 3 term in \rf{4.17} 
which  at present  is   not known beyond the leading order in the curvature. 

The explicit form  of $Q$ is found  to be 
\begin{align}
\la{D.2}
Q = &- \tfrac{8}{3} R_{b}{}^{c} R_{c}{}^{d} \eps_{ad} \nabla^{b}h^{a} + 
\tfrac{8}{3} R_{a}{}^{c} R_{c}{}^{d} \eps_{bd} \nabla^{b}h^{a} + 8 \
R_{b}{}^{e}{}_{c}{}^{f} C_{adef} \eps^{cd} \nabla^{b}h^{a} \
\nonumber \\ 
& - 8 R_{b}{}^{e}{}_{c}{}^{f} C_{afde} \eps^{cd} 
\nabla^{b}h^{a} - 8 R_{a}{}^{e}{}_{c}{}^{f} C_{bdef} \
\eps^{cd} \nabla^{b}h^{a} + 8 R_{a}{}^{e}{}_{c}{}^{f} \
C_{bfde} \eps^{cd} \nabla^{b}h^{a} \nonumber \\ 
& + \tfrac{8}{3} R_{b}{}^{c} R_{adce} \eps^{de} \nabla^{b}h^{a} -  
\tfrac{8}{3} R_{a}{}^{c} R_{bdce} \eps^{de} \nabla^{b}h^{a} + 
\tfrac{8}{3} R_{b}{}^{c} \nabla^{b}h^{a} \nabla_{d}
\nabla_{c}\eps_{a}{}^{d} \nonumber \\ 
& -  \tfrac{8}{3} R_{a}{}^{c} \nabla^{b}h^{a} \nabla_{d}
\nabla_{c}\eps_{b}{}^{d} - 8 \nabla_{a}C_{bcde} 
\nabla^{b}h^{a} \nabla^{e}\eps^{cd} + 8 \nabla_{b}C_{acde} 
\nabla^{b}h^{a} \nabla^{e}\eps^{cd} \nonumber \\ 
& + 8 R_{bcde} \nabla^{b}h^{a} \nabla^{e}\nabla_{a}\eps^{cd} - 8 \
C_{bcde} \nabla^{b}h^{a} \nabla^{e}\nabla_{a}\eps^{cd} - 8 \
R_{acde} \nabla^{b}h^{a} \nabla^{e}\nabla_{b}\eps^{cd} \nonumber \\ 
& + 8 C_{acde} \nabla^{b}h^{a} \nabla^{e}
\nabla_{b}\eps^{cd} + \tfrac{16}{3} R_{abde} \nabla^{b}h^{a} \nabla^{e}\
\nabla_{c}\eps^{cd} + \tfrac{8}{3} R_{adbe} \nabla^{b}h^{a} 
\nabla^{e}\nabla_{c}\eps^{cd} \nonumber \\ 
& -  \tfrac{8}{3} R_{aebd} \nabla^{b}h^{a} \nabla^{e}
\nabla_{c}\eps^{cd} + 8 R_{bcde} \nabla^{b}h^{a} \nabla^{e}
\nabla^{d}\eps_{a}{}^{c} + 8 R_{bdce} \nabla^{b}h^{a} \nabla^{e}
\nabla^{d}\eps_{a}{}^{c} \nonumber \\ 
& - 8 R_{acde} \nabla^{b}h^{a} \nabla^{e}\nabla^{d}\eps_{b}{}^{c} - \
8 R_{adce} \nabla^{b}h^{a} \nabla^{e}\nabla^{d}\eps_{b}{}^{c}.
\end{align}
Integrating by parts to remove the covariant derivatives from $\eps_{ab}$ and using the Einstein-space  curvature identities 
we arrive at 
\begin{align}
\la{D.3}
Q =&- \tfrac{4}{3} R \eps_{bc} \nabla_{a}\nabla^{c}\nabla^{b}h^{a} + 
\tfrac{4}{3} R \eps_{ac} \nabla_{b}\nabla^{c}\nabla^{b}h^{a} + 
\tfrac{1}{6} R^2 \eps_{ab} \nabla^{b}h^{a} \nonumber \\ 
& + \tfrac{2}{3} R \eps_{bc} \nabla^{c}\nabla_{a}\nabla^{b}h^{a} -  \
\tfrac{4}{3} R \eps_{ac} \nabla^{c}\nabla_{b}\nabla^{b}h^{a} - 8 \
\eps^{de} \nabla_{a}C_{bdce} \nabla^{c}\nabla^{b}h^{a} 
\nonumber \\ 
& + 8 \eps^{de} \nabla_{b}C_{adce} \nabla^{c}
\nabla^{b}h^{a} + \tfrac{2}{3} R \eps_{bc} \nabla^{c}\nabla^{b}
\nabla_{a}h^{a} - 8 C_{bdce} \eps_{a}{}^{e} \nabla^{d}
\nabla^{c}\nabla^{b}h^{a} \nonumber \\ 
& - 8 C_{becd} \eps_{a}{}^{e} \nabla^{d}\nabla^{c}
\nabla^{b}h^{a} + 8 C_{adce} \eps_{b}{}^{e} \nabla^{d}
\nabla^{c}\nabla^{b}h^{a} + 8 C_{aecd} \eps_{b}{}^{e} 
\nabla^{d}\nabla^{c}\nabla^{b}h^{a} \nonumber \\ 
& -  \tfrac{16}{3} C_{abce} \eps_{d}{}^{e} \nabla^{d}
\nabla^{c}\nabla^{b}h^{a} -  \tfrac{8}{3} C_{acbe} \
\eps_{d}{}^{e} \nabla^{d}\nabla^{c}\nabla^{b}h^{a} +
 \tfrac{8}{3} C_{aebc} \eps_{d}{}^{e} \nabla^{d}
\nabla^{c}\nabla^{b}h^{a}\notag \\
& -  \tfrac{16}{3} \eps^{de} \nabla^{c}
\nabla^{b}h^{a} \nabla_{e}C_{abcd}  -  \tfrac{8}{3} \eps^{de} \nabla^{c}\nabla^{b}h^{a} 
\nabla_{e}C_{acbd} + \tfrac{8}{3} \eps^{de} \nabla^{c}
\nabla^{b}h^{a} \nabla_{e}C_{adbc}.
\end{align}
Symmetrizing the covariant derivatives and using the Bianchi identities   gives  
\begin{align}
\la{D.4}
Q &= \tfrac{4}{3} R_{b}{}^{(de)f} R_{cdef} \eps_{a}{}^{c} \nabla^{[b}h^{a]}  
\ .
\end{align}
This vanishes after using $\eps_{ab} = \eps_{ba}$   and  the  identities  valid in
 Einstein space  ({\em cf.}  (\ref{A.1}))
\be
\la{D.5}
R^{acde} R^{b}{}_{cde} =  \tfrac{1}{4} g^{ab} R_{cdef} R^{cdef},
\qquad \qquad 
R^{acde} R^{b}{}_{dce} =  \tfrac{1}{4} g^{ab} R_{cedf} R^{cdef}.
\ee

\section{Restricted form of   spin 3  gauge invariance } 
\la{app:basis}

If we consider the spin 3 gauge   transformations 
 with the      gauge parameter constrained by  $\nabla^{a}\eps_{ab}=0$
 then  the  transformations in (\ref{4.16})  satisfy 
\be
\la{E.1}
\delta' h_{a} = -6\,\nabla^{b}\,\nabla^{c}\,\delta' h_{abc} \ , \ \ \ \ \ \ \qquad \ \ \ \   \delta' \equiv  \delta\Big|_{ \nabla^{a}\eps_{ab}=0  }  \ . 
\ee
This means that we can introduce a  new  spin-1 field $\hA_{a}$
which will be  neutral with respect to the  restricted spin 3   gauge   transformation 
\be
\la{E.2}
\hA_{a} = h_{a}+6\,\nabla^{b}\nabla^{c}\,h_{abc} \ , \ \ \ \ \ \ \ \ \   \qquad \delta' \hA_a =0 \ . 
\ee
Then the  spin 1 and 3  interaction terms in  \rf{2.21}    may be written as 
\be
\la{E.3}
h^{a}\,J_{a}+h^{abc}\,J_{abc} = \hA^{a}\,J_{a}+h^{abc}\,\widetilde J_{abc},\qquad \qquad 
\widetilde J_{abc} = J_{abc}- 6 \nabla_{(a}\,\nabla_{b}\,J_{c)} \ , 
\ee
where $\widetilde J_{abc} $  is thus the same as in \rf{2.24}, {\em i.e.} 
\be
\la{E.4}
h^{abc}\,\widetilde J_{abc} = 60\,i\,h^{abc}\,\nabla_{a}\nabla_{b}\ovp\ \nabla_{c}\vp  + c.c. \ . 
\ee
This term  is   thus     invariant  under    the   restricted gauge transformations    in  an  Einstein  background. 

 The quadratic  part of the induced action is again 
of the form (\ref{4.16}), but  now    written in terms of the new  vector  field $\hA_{a}$ (with field strength $\hF $) 
\begin{align}
\la{E.5}
\text{S}^{(2)} = & \int d^{4}x\,\sqrt{g}\,\Big[\te -\frac{1}{12}\,\hF^{2}_{ab}+{\mathscr{\widetilde  L}_{13}}
+h_{(3)}\, \widetilde {\mc  O}_{6}\, h_{(3)}\Big] \ , \\
\la{E.6}
 {\mathscr { \widetilde L}_{13}} =  & \hA^{a}\,\langle J_{a}\,\widetilde J_{bcd}\rangle_{_{\text{UV}}}  h^{bcd}
 = \hF^{ab}   \big(2\,\nabla_a \,\nabla_c \,\nabla_d\, h\indices{_{b}^{cd}} 
+ 8\, C_{acdp} \,\nabla^p\, h\indices{_{b}^{cd}} \big)
 \  .
\end{align}
Note that in an Einstein background  there is no nontrivial $\hA_{(1)} h_{(3)}$ tadpole contributions   so \rf{E.6}  is automatically  vector gauge invariant. 

Since $\delta' \hA_a=0$, 
 the $h_{(3)}\,  \widetilde {\mc  O}_6 \, h_{(3)}$ term in (\ref{E.5})
that     should be   equal to  $h_{(3)}\langle \widetilde  J\ \widetilde J\rangle_{_{\rm UV}} h_{(3)}$
(up to a possible  tadpole contribution) 
 should  be   invariant   under the restricted spin 3   gauge transformations on its own.

It is easy to see   
  why  (\ref{E.6}) is  spin 3  gauge invariant   using  (\ref{E.1}):
\begin{align}
\no 
\delta' \mathscr L_{13} = &\hF^{ab} \big(2\,\nabla_a \,\nabla_c \,\nabla_d\, \delta' h\indices{_{b}^{cd}} 
+ 8\, C_{acdp} \,\nabla^p\, \delta' h\indices{_{b}^{cd}}    \big) 
=\hF^{ab} \big(-\tfrac{1}{3}\,\nabla_a \,\delta' h_{b}  + 8\, C_{acdp} \,\nabla^p\, \delta' h\indices{_{b}^{cd}}  \big) \\
=&\delta' \big(\te -\frac{1}{12}\,F_{ab}^{2} +    8\,  F^{ab} \, C_{acdp} \,\nabla^p\,  h\indices{_{b}^{cd}}     \big)\Big|_{h_{(1)}\eps} =0  \ . 
\end{align}
Here we used  that 
  the two terms   in the second line  are as in \rf{4.17}  and    thus are invariant under the spin 3  transformations 
 modulo   $O( h_{(3)}\eps)$ term. 

\section{Some  expressions used in  section   \ref{sec:NT} }
\la{app:X}

Here  we   present  the  relations   used  in eqs. \rf{4.20},\rf{4.21},\rf{4.22}.
 The  tensor $\mc V_{ab}$ in the $h_{(2)} h_{(3)} \ovp \vp$  vertex in (\ref{4.21}) is  given by 
\begin{align}
\la{F.1}
\mc V_{ab} = &3\,i\, (
 -4\,\partial_{c}h_{abd}\,\partial^{d}\ovp\,\partial^{c}\vp
+ 4\,
\partial_{d}h_{abc}\,\partial^{d}\ovp\,\partial^{c}\vp
+ 10\,h_{bcd}\,
\partial^{c}\ovp\,\partial_{a}{}^{d}\vp
 -10\,h_{bcd}\,\partial_{a}{}^{d}
\ovp\,\partial^{c}\vp \notag \\
&  + 10\,h_{acd}\,\partial^{c}\ovp\,\partial_{b}
{}^{d}\vp
 -10\,h_{acd}\,\partial_{b}{}^{d}\ovp\,\partial^{c}\vp
 -4\
\,h_{abd}\,\partial^{c}\ovp\,\partial_{c}{}^{d}\vp
+ 4\,h_{abd}\,
\partial_{c}{}^{d}\ovp\,\partial^{c}\vp\notag \\
 &  -7\,\partial_{a}h_{bcd}\,\ovp\
\,\partial^{cd}\vp
 -7\,\partial_{b}h_{acd}\,\ovp\,\partial^{cd}\vp
+ 12\,\partial_{d}h_{abc}\,\ovp\,\partial^{cd}\vp
+ 7\,\partial_{a}\
h_{bcd}\,\partial^{cd}\ovp\,\vp +7\,\partial_{b}h_{acd}\,
\partial^{cd}\ovp\,\vp\notag \\
&  -12\,\partial_{d}h_{abc}\,\partial^{cd}\ovp\,\vp
+ 3\,\
h_{abc}\,\partial^{c}\ovp\,\partial^{d}{}_{d}\vp
 -3\,h_{abc}\,
\partial^{d}{}_{d}\ovp\,\partial^{c}\vp
+ 7\,\partial^{d}{}_{d}h_{abc}\,\
\ovp\,\partial^{c}\vp\notag \\
&  -7\,\partial^{d}{}_{d}h_{abc}\,\partial^{c}
\ovp\,\vp
+ 12 \
\eta_{bc}\,h_{ade}\,\partial^{c}\ovp\,\partial^{de}\vp
+ 12 \eta_{ac}\
\,h_{bde}\,\partial^{c}\ovp\,\partial^{de}\vp
 -12 \
\eta_{bc}\,h_{ade}\,\partial^{de}\ovp\,\partial^{c}\vp\notag \\
&  -12 \
\eta_{ac}\,h_{bde}\,\partial^{de}\ovp\,\partial^{c}\vp
 -8\,h_{b}
{}^{cd}\,\ovp\,\partial_{acd}\vp
+ 8\,h_{b}{}^{cd}\,\partial_{acd}\ovp\,\vp
 -8\,h_{a}{}^{cd}\,\ovp\,\partial_{bcd}\vp
+ 8\,h_{a}{}^{cd}\,
\partial_{bcd}\ovp\,\vp\notag \\
& +\,h_{ab}{}^{c}\,\ovp\,\partial_{c}{}^{d}{}_{d}\vp
 -\,h_{ab}{}^{c}\,\partial_{c}{}^{d}{}_{d}\ovp\,\vp)\ . 
\end{align}
The contribution $\mathscr L_{233}^{(a)}$ in \rf{4.20}   coming 
 from the triangle diagram (a) in Fig. \ref{fig:233} is
\begin{align}
 \la{F.2}
\!\! \mathscr L_{233}^{(a)} = &-\tfrac{15}{7}\,h^{ab}\,(
 -42 \Box h_{a}{}^{cd} \Box^{2} h_{bcd} + 35 \partial_{a}\Box h^{cde} \
\partial_{b}\Box h_{cde} + 42 \partial_{a}h^{cde} \
\partial_{b}\Box^{2} h_{cde}  \nonumber \\ 
& + 54 \partial_{a}h^{cde} \
\partial_{e}\Box^{2} h_{bcd}  + 18 \partial_{b}\Box^{2} h_{cde} \partial^{e}h_{a}{}^{cd} - 108 \
\partial_{d}\Box^{2} h_{bce} \partial^{e}h_{a}{}^{cd} - 54 \
\partial_{e}\Box^{2} h_{bcd} \partial^{e}h_{a}{}^{cd} \nonumber \\ 
& + 54 \partial_{b}\Box h_{cdf} \partial^{f}\Box h_{a}{}^{cd} - 102 \
\partial_{d}\Box h_{bcf} \partial^{f}\Box h_{a}{}^{cd} - 51 \
\partial_{f}\Box h_{bcd} \partial^{f}\Box h_{a}{}^{cd} + 42 \Box \
h^{cde} \partial_{ab}\Box h_{cde} \nonumber \\ 
& + 24 h^{cde} \partial_{ab}\Box^{2} h_{cde} + 28 \Box^{2} h_{cde} \
\partial_{ab}h^{cde} - 12 \Box h^{cde} \partial_{be}\Box h_{acd} + \
264 \partial_{a}{}^{f}h^{cde} \partial_{be}\Box h_{cdf} \nonumber \\ 
& - 12 \Box^{2} h_{cde} \partial_{b}{}^{e}h_{a}{}^{cd} + 88 \
\partial_{a}{}^{f}h^{cde} \partial_{bf}\Box h_{cde} + 72 \Box h^{cde} \
\partial_{de}\Box h_{abc} + 72 \partial_{a}{}^{f}h^{cde} \
\partial_{de}\Box h_{bcf} \nonumber \\ 
& + 54 h^{cde} \partial_{de}\Box^{2} h_{abc} + 30 \Box^{2} h_{cde} \
\partial^{de}h_{ab}{}^{c} + 72 \partial_{a}{}^{f}h^{cde} \
\partial_{ef}\Box h_{bcd} - 18 h_{a}{}^{cd} \
\partial^{e}{}_{e}\Box^{2} h_{bcd} \nonumber \\ 
& + 48 \partial_{bd}\Box h_{cef} \partial^{ef}h_{a}{}^{cd} + 48 \
\partial_{bf}\Box h_{cde} \partial^{ef}h_{a}{}^{cd} - 72 \
\partial_{cd}\Box h_{bef} \partial^{ef}h_{a}{}^{cd} 
- 288 \partial_{df}\Box h_{bce} \partial^{ef}h_{a}{}^{cd} \notag \\
& - 72 \
\partial_{ef}\Box h_{bcd} \partial^{ef}h_{a}{}^{cd} + 144 \
\partial^{f}h^{cde} \partial_{abe}\Box h_{cdf}  + 48 \partial^{f}h^{cde} \partial_{abf}\Box h_{cde}
 + 150 \
\partial_{e}\Box h_{cdf} \partial_{ab}{}^{f}h^{cde}\notag \\
& + 50 \partial_{f}\
\Box h_{cde} \partial_{ab}{}^{f}h^{cde}  + 90 \partial_{a}{}^{fh}h^{cde} \partial_{bde}h_{cfh} + 180 \
\partial_{a}{}^{fh}h^{cde} \partial_{beh}h_{cdf} - 12 \
\partial_{d}\Box h_{cef} \partial_{b}{}^{ef}h_{a}{}^{cd} \nonumber \\ \
& - 12 \partial_{f}\Box h_{cde} \partial_{b}{}^{ef}h_{a}{}^{cd} + 30 \
\partial_{a}{}^{fh}h^{cde} \partial_{bfh}h_{cde} + 36 \
\partial^{f}h^{cde} \partial_{cde}\Box h_{abf}  + 108 \partial^{f}h^{cde} \partial_{def}\Box h_{abc}
\notag \\
& + 30 \
\partial_{c}\Box h_{def} \partial^{def}h_{ab}{}^{c} + 90 \partial_{f}\
\Box h_{cde} \partial^{def}h_{ab}{}^{c} + 12 \partial_{bcd}h_{efh} \partial^{efh}h_{a}{}^{cd} + 72 \
\partial_{bdh}h_{cef} \partial^{efh}h_{a}{}^{cd} \notag \\
&+ 36 \
\partial_{bfh}h_{cde} \partial^{efh}h_{a}{}^{cd}  - 54 \partial_{cdh}h_{bef} \partial^{efh}h_{a}{}^{cd} - 108 \
\partial_{dfh}h_{bce} \partial^{efh}h_{a}{}^{cd} - 18 \
\partial_{efh}h_{bcd} \partial^{efh}h_{a}{}^{cd} \nonumber \\ 
& + 96 \partial^{fh}h^{cde} \partial_{abde}h_{cfh} + 192 \
\partial^{fh}h^{cde} \partial_{abeh}h_{cdf} + 32 \partial^{fh}h^{cde} \
\partial_{abfh}h_{cde} + 48 \partial^{fh}h^{cde} \
\partial_{cdeh}h_{abf} \nonumber \\ 
& + 72 \partial^{fh}h^{cde} \partial_{defh}h_{abc}
 ).
 \end{align}
 The  contribution 
 $\mathscr L_{233}^{(b)}$ in \rf{4.20}   coming 
 from the  bubble diagram (b) in Fig. \ref{fig:233} is  much simpler 
 \begin{align}
 \la{F.3}
 \mathscr L_{233}^{(b)} &=
 - \tfrac{270}{7}\, h^{ab}\, h_{a}{}^{cd}\Box^{3} h_{bcd} \ .
 \end{align}

\bibliography{BT-Biblio}
\bibliographystyle{JHEP}

\end{document}